\pdfoutput=1

\documentclass[11pt]{article}

\usepackage[final]{acl}
\usepackage{multirow}
\usepackage[table,xcdraw]{xcolor}
\usepackage{times}
\usepackage{titlesec}
\usepackage{latexsym}
\usepackage{bbding}
\usepackage{makecell}
\usepackage{enumitem}
\usepackage{graphicx}
\usepackage{booktabs}
\usepackage{array}
\usepackage{longtable}
\usepackage{lipsum}
\usepackage{array}
\usepackage{multirow}
\usepackage{graphicx}
\usepackage{longtable}
\usepackage{hyperref}
\usepackage{times}
\usepackage{svg}
\usepackage{latexsym}
\usepackage{multirow}
\usepackage{threeparttable}

\usepackage{amsmath}
\usepackage{graphicx}
\usepackage{booktabs}
\usepackage{adjustbox}
\usepackage[T1]{fontenc}
\usepackage{algorithm,algorithmic}
\usepackage[utf8]{inputenc}
\usepackage{multirow}
\usepackage[table,xcdraw]{xcolor}
\usepackage{microtype}

\usepackage{inconsolata}

\setlist{nosep}

\usepackage{amsmath}
\title{MA-GTS: A Multi-Agent Framework for Solving Complex Graph Problems in Real-World Applications}

\author{
 \textbf{Zike Yuan\textsuperscript{1,2}},
 \textbf{Ming Liu\textsuperscript{1,2,*}},
 \textbf{Hui Wang\textsuperscript{2,*}},
 \textbf{Bing Qin\textsuperscript{1,2,*}}
\\
 \textsuperscript{1}Harbin Institute of Technology, Shenzhen, China,\\
 \textsuperscript{2}Peng Cheng Laboratory, Shenzhen, China\\
 \texttt{\{yuanzk,wangh06\}@pcl.ac.cn}\\
 \texttt{\{mliu,qinb\}@ir.hit.edu.cn}
\\
}

\begin{document}
\maketitle
\begin{abstract}
Graph-theoretic problems arise in real-world applications like logistics, communication networks, and traffic optimization. These problems are often complex, noisy, and irregular, posing challenges for traditional algorithms. Large language models offer potential solutions but face several challenges, including limited accuracy, input length constraints, and suboptimal algorithm selection. To address these challenges, we propose \textbf{MA-GTS} (\textbf{M}ulti-\textbf{A}gent \textbf{G}raph \textbf{T}heory \textbf{S}olver), a multi-agent framework that decomposes these complex problems through agent collaboration. MA-GTS maps the implicitly expressed text-based graph data into clear, structured graph representations and dynamically selects the most suitable algorithm based on problem constraints and graph structure scale. 
We validate MA-GTS using the \textbf{G-REAL} dataset, a real-world-inspired graph theory dataset we created. Experimental results show that MA-GTS outperforms state-of-the-art methods in cost-effectiveness, accuracy, and scalability, achieving strong results on multiple benchmarks (G-REAL \textbf{93.6\%}, GraCoRe \textbf{96.9\%} NLGraph \textbf{98.4\%}) with robust performance on both closed- and open-source base models. MA-GTS and G-REAL are open-sourced at \href{https://github.com/ZIKEYUAN/MA-GTS.git}{https://github.com/ZIKEYUAN/MA-GTS.git}.
\end{abstract}
\renewcommand{\thefootnote}{}
\footnotetext{*  B. Qin, H. Wang and M. Liu are corresponding authors.}
\section{Introduction}
Graph-theoretic problems have extensive applications in domains such as logistics scheduling, communication networks, production planning, and traffic optimization \cite{suvery}. These problems typically involve a large number of nodes and edges, coupled with complex constraints and dynamic variations, making their solution highly challenging \cite{bondy2008graph}. Despite significant advancements in graph theory and algorithmic design, traditional approaches remain computationally expensive and inefficient when handling large-scale, high-complexity problems. Existing methods, including exact algorithms, greedy strategies, and dynamic programming \cite{bellman1966dynamic}, perform well on small-scale instances. However, as problem size increases, their computational complexity and memory requirements grow exponentially, rendering them impractical for real-world applications. While heuristic methods 
\cite{kokash2005introduction} can improve performance under specific conditions, they often suffer from local optima and require extensive parameter tuning and model selection. Therefore, developing efficient and scalable solution frameworks capable of addressing the computational demands and structural variability of complex graph-theoretic problems remains a critical research challenge.
\begin{figure}[t]
  \includegraphics[width=\columnwidth]{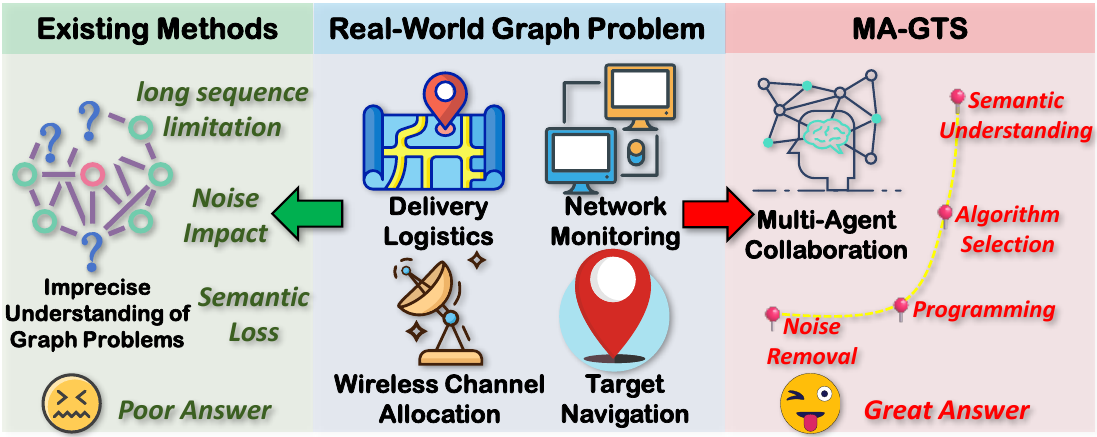}
  \vspace{-20pt}
  \caption{MA-GTS leverages multi-agent collaboration to overcome noise and semantic loss in real-world graph problems, leading to better answers.}
     \vspace{-23pt}
  \label{fig:0}
\end{figure}

\begin{figure*}[htbp]
\centering
  \includegraphics[width=0.9\textwidth]{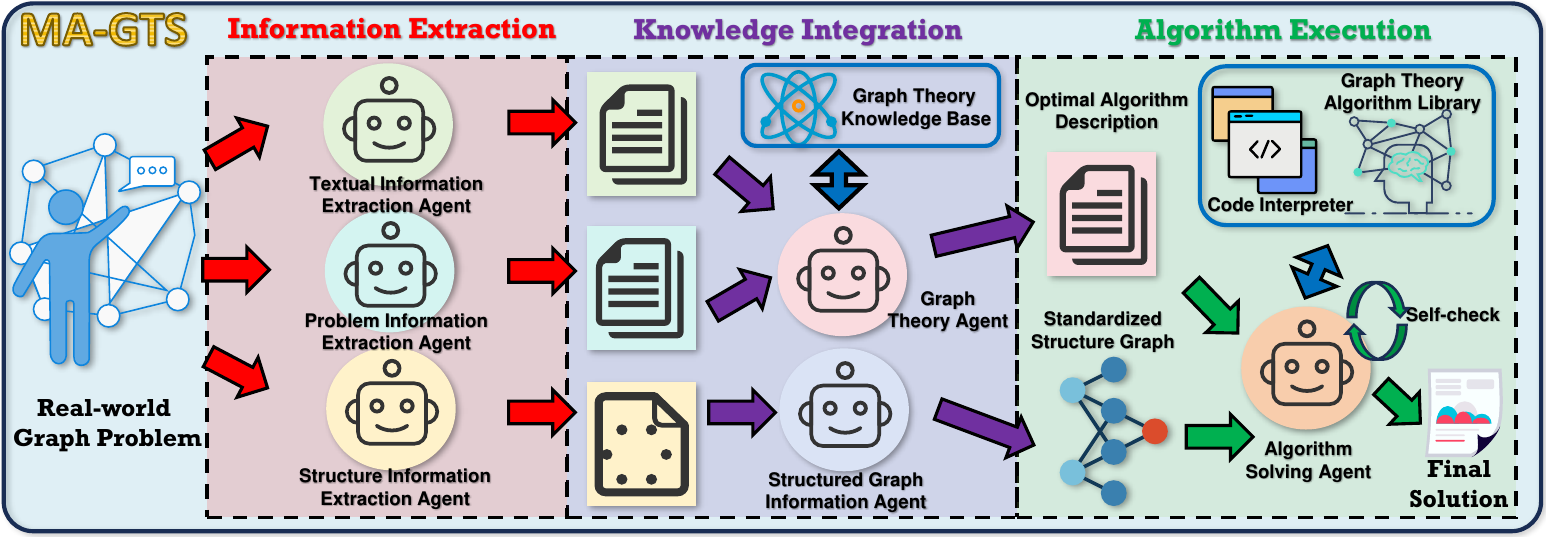}
  \vspace{-8pt}
  \caption{MA-GTS framework for solving real-world graph problems, consisting of three layers: Information Extraction, Knowledge Integration, and Algorithm Execution, each with specialized agents.}
  \label{fig:1}
  \vspace{-15pt}
\end{figure*}

Recent advancements in LLMs have spurred interest in their applications for graph-theoretic problems. Leveraging their natural language processing (NLP) capabilities, LLMs can serve as  \textbf{scene interpreters} (mapping real-world problems to graph models),  \textbf{graph extractors} (identifying graph structures from unstructured data), and  \textbf{graph algorithm invokers} (assisting in solving and optimizing graph-based problems), addressing certain limitations of traditional algorithms. However, significant challenges remain in existing methods (LLMs and simple multi-agent framework). Figure \ref{fig:0} clearly illustrates the challenges existing methods face when addressing real-world graph problems. \textbf{Firstly}, LLMs rely on statistical pattern matching rather than strict mathematical computations, limiting their reasoning accuracy and making them unreliable for NP-hard problems \cite{hochba1997approximation}.  \textbf{Secondly}, their ability to handle large-scale graphs is limited by the Transformer \cite{vaswani2017attention} architecture's context window and computational complexity, which restricts their capacity to capture global information. \textbf{Finally}, LLMs lack the ability to decompose and map real-world graph theory problems, which often contain complex textual noise and implicit graph structures. In summary, existing methods struggle to effectively handle long texts and graph problems in real-world scenarios.  Problems like disordered nodes, noisy text, and poor algorithm choices can all affect the quality of graph modeling, text understanding, and the interpretability of reasoning. These limitations highlight the inadequacy of existing methods for solving complex graph-theoretic problems in real-world applications and underscore the need for more efficient and scalable paradigms.

To tackle these challenges, we propose \textbf{MA-GTS}(\textbf{M}ulti-\textbf{A}gent \textbf{G}raph \textbf{T}heory \textbf{S}olver), an innovative multi-agent framework designed to address complex real world graph-theoretic problems through agent collaboration and competition. Figure \ref{fig:1} illustrates the framework, which incorporates a multi-agent coordination mechanism allowing agents to perform local searches independently while sharing information and cooperating, thus improving solution efficiency and accuracy. MA-GTS analyzes the original real-world problem textual data, filters out noise, and extracts key graph data and problem-specific details, reducing the text length that LLMs must process and enhancing reasoning efficiency. MA-GTS selects the optimal graph algorithm based on refined text and adjusts the graph’s textual representation to match the algorithm, improving reasoning and solution quality. This coordination mitigates the limitations of LLMs in implicit graph structure modeling, ensuring efficient solutions for complex graph tasks. Additionally, dynamic agent interactions enable the framework to address large-scale problems and adapt to complex constraints and dynamic changes.
\begin{figure*}[htbp]
\centering
  \includegraphics[width=1\textwidth]{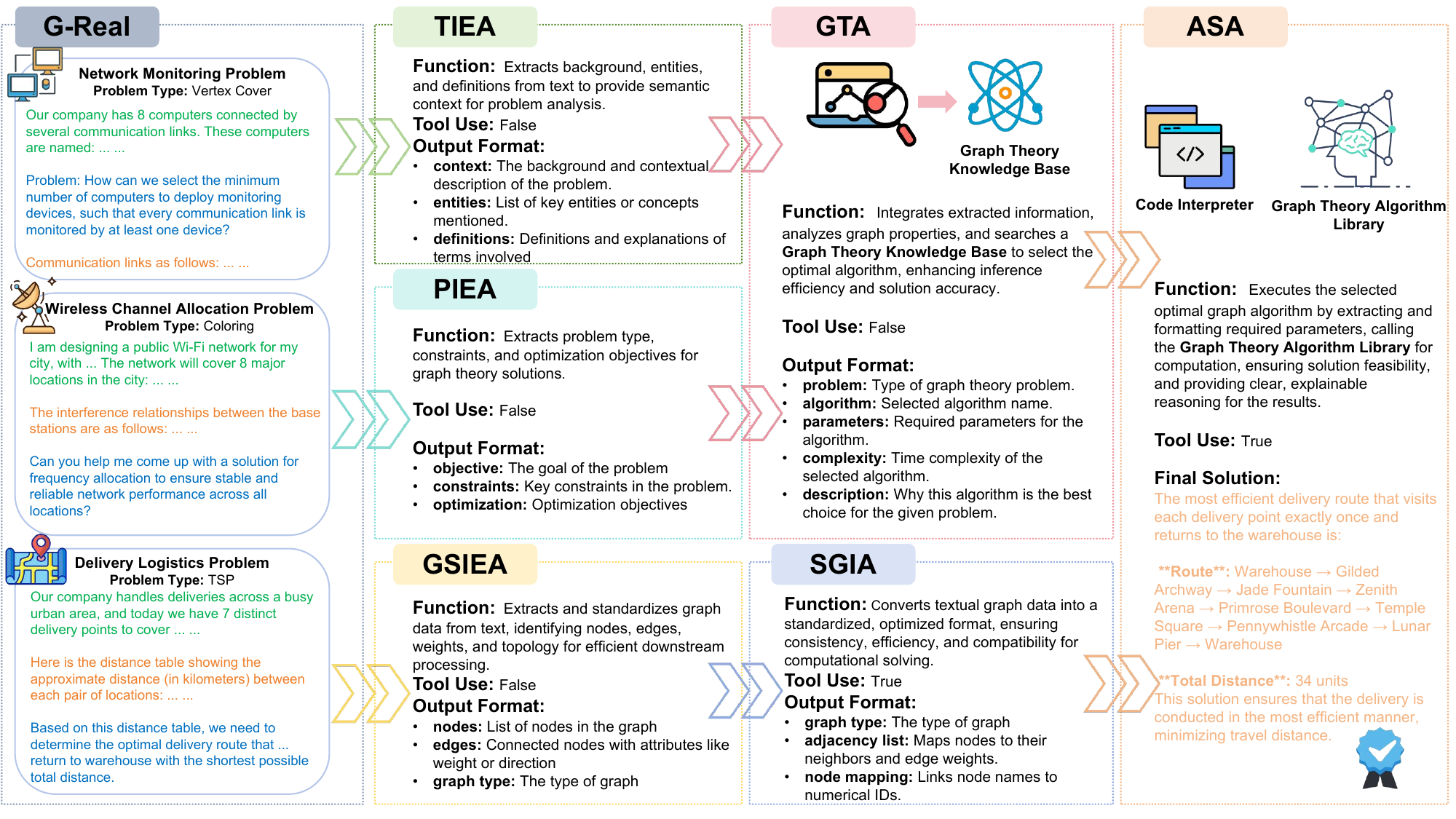}
  \vspace{-20pt}
  \caption{This figure details the G-REAL dataset's composition and features, along with the full MA-GTS graph problem-solving pipeline, outlining each component's functions and input/output formats.}
  \label{fig:2}
  \vspace{-15pt}
\end{figure*}

\vspace{-10pt}
To validate the effectiveness of the multi-agent framework, we introduce the \textbf{G-REAL} dataset, designed to simulate complex graph theory problems relevant to real-world scenarios. Unlike traditional datasets that rely on simple textual descriptions of graph structures, G-REAL better reflects practical applications for large-scale models. Experiments comparing MA-GTS with state-of-the-art open-source and closed-source LLMs (including three closed-source and three open-source models), as well as with a general multi-agent framework and a graph-specific multi-agent framework, show that MA-GTS significantly outperforms existing LLMs and multi-agent frameworks in terms of efficiency and accuracy, under both direct reasoning and Chain of Thought (CoT) \cite{wei2022chain} reasoning settings. Notably, it excels in solving large-scale problems with complex constraints, offering superior scalability, robustness, and cost-effectiveness. The primary contributions of this study are as follows:
\begin{itemize}[left=0pt]
\item First, we propose an innovative multi-agent framework, MA-GTS, which overcomes the limitations of traditional graph theory algorithms in large-scale complex problems, achieving state-of-the-art performance in our tests.
\item Second, we constructed a real-world graph theory dataset, G-REAL, that aligns with practical needs, providing the necessary data support for validating the effectiveness of the algorithm.
\item Finally, by introducing novel collaboration mechanisms and strategies, we achieve efficient and precise graph theory problem-solving within the multi-agent system, demonstrating its substantial potential in real-world application scenarios.
\end{itemize}

\section{Related Work}
\textbf{LLMs for Graph}: Recent advancements in LLMs for graph tasks have led to significant contributions in methodology and evaluation. These tasks are often classified into Enhancer, Predictor, and Alignment types \cite{suvery}. Notably, \cite{2} presents a roadmap for unifying LLMs with Knowledge Graphs (KGs), while \cite{3} proposes an end-to-end method for solving graph-related problems, \cite{cao2024graphinsight} improves LLMs' understanding of graph structures by addressing positional biases and incorporating an external knowledge base. On the evaluation front, several benchmarks have been introduced. NLGraph \cite{NLgraph-6} offers a simple test dataset for graph tasks, and GPT4Graph \cite{gpt4graph-7} evaluates LLM capabilities on semantic tasks. GraCoRe \cite{yuan2025gracore} comprehensively verifies the graph understanding and reasoning capabilities of LLM. In addition to these representative benchmarks, ProGraph \cite{li2024can}, GraphArena \cite{tang2024grapharena}, GLBench \cite{li2024glbench}, etc. are also widely used. Other notable works include \citep{8}, which assesses LLMs in graph data analysis, and \citep{10}, which designs a hint method for graph tasks.

\noindent\textbf{LLM Agents}: Several multi-agent frameworks have been proposed to improve coordination and efficiency in complex tasks. MetaGPT \cite{hong2023metagpt} embeds human workflows into LLMs to reduce hallucinations. CAMEL \cite{li2023camel} enables autonomous agent cooperation aligned with human goals, and its extension OWL \cite{owl2025} builds on this. AutoGen \cite{wu2023autogen} offers a flexible framework for customizing agent interactions via natural language and code. Additionally, \cite{li2024graphteam} addresses simple graph problems. Multi-agent frameworks like GraphTeam \cite{li2024graphteam}, GCoder \cite{zhang2024gcoder}, and GraphAgent \cite{hu2024scalable} can enhance the reasoning ability of LLMs through multiple interactions, but they are mainly applied to standard graph structures, and their effectiveness on real-world graph theory problems remains uncertain.
\section{MA-GTS}
We consider a real-world graph problem \( P \), modeled as a graph \( G\). The system uses a graph theory knowledge base \( \mathcal{K}_{\mathcal{G}} \) and an algorithm library \( \mathcal{L}_{\text{code}} \) to support problem understanding and solving. From \( P \), it extracts the textual description \( \mathcal{T} \), identifies the problem type \( \mathcal{P} \), and constructs the graph structure \( G \). The objective is to automatically select an appropriate algorithm \( Alg^* \), apply it to the structured graph \( G' \), and iteratively optimize the solution \( S^n \).

The MA-GTS framework adopts a hierarchical processing paradigm, comprising three layers: the \textbf{Information Extraction Layer(IEL)}, the \textbf{Knowledge Integration Layer(KIL)}, and the \textbf{Algorithm Execution Layer(AEL)}. These layers interact through a hierarchical collaborative communication mechanism, enabling an end-to-end pipeline that processes unstructured data and solves complex graph-theoretic problems. Additionally, to support the knowledge base of MA-GTS, we have constructed the Graph Theory Knowledge Base and Graph Theory Algorithm Library. More information about them in the Appendix \ref{sec:A}.

The IEL processes text and structured data to extract graph information and identify problem types for standardized input. The KIL builds structured graph data using graph theory and optimization to enhance accuracy and scalability. The AEL runs specified algorithms and performs self-checks to efficiently solve complex graph problems. Figure \ref{fig:2} shows each agent’s function by layer.

By leveraging agent collaboration, MA-GTS ensures efficient problem-solving, high scalability, and adaptability to complex constraints, offering a novel solution for real-world graph-theoretic challenges. The specific functionalities of each agent are detailed as follows:
\subsection{Information Extraction Layer (IEL)}
The IEL extracts relevant information from text and unstructured data, structures it for downstream use, and filters out irrelevant content to sharpen problem-specific details. It also captures implicit graph structures to boost efficiency and reduce the effects of text length on LLMs inference.

\noindent\textbf{Textual Information Extraction Agent (TIEA)}: The TIEA analyzes real-world graph problems to extract key textual information unrelated to graph structure or solution goals. Using NLP, it identifies and structures context, background, entities, concepts, and definitions, organizing semantic content to support later analysis. The output is standardized for downstream processing.
\renewcommand{\algorithmicrequire}{\textbf{Input:}}
\renewcommand{\algorithmicensure}{\textbf{Output:}}
\begin{algorithm}[t]
\footnotesize
\caption{Pipeline of MA-GTS}
\begin{algorithmic}[1]

\REQUIRE Real-world graph problem $P$, graph theory knowledge base $\mathcal{K_G}$, graph theory algorithm library $\mathcal{L}_{code}$., self check number $N_{check}$
   
\ENSURE Optimized solution $S^n$
\STATE \textbf{Step 1: Information Extraction Layer}
\STATE Extract textual information: $\mathcal{T} \gets A_{TIEA}(P)$
\STATE Identify problem type: $\mathcal{P} \gets A_{PIEA}(P)$
\STATE Extract graph structure: $G \gets  A_{GSIEA}(P)$\;
\STATE Generate extracted information set: $(\mathcal{T}, \mathcal{P}, G)$

\STATE \textbf{Step 2: Knowledge Integration Layer}
\STATE Select best algorithm: 
\STATE \hspace{0.2cm} $\mathcal{L}_{\mathcal{P}} \gets  A_{GTA}(\mathcal{T},\mathcal{P}, \mathcal{K_G})$
\STATE  \hspace{0.2cm} $Alg^* \gets \arg\operatorname{opt}_{Alg_i \in \mathcal{L}_{\mathcal{P}}}  A_{GTA}(Alg_i, \mathcal{T})$
\STATE Get structured graph: $G' \gets A_{SGIA}(G)$
\STATE Define structured problem: $(G', Alg^*)$
\STATE \textbf{Step 3: Algorithm Execution Layer}
\STATE Load algorithm code: 
\STATE \hspace{0.2cm} $Code_{Alg^*} \gets A_{ASA}(Alg^*, \mathcal{L}_{code})$
\STATE Get algorithm output : 
\STATE \hspace{0.2cm} $S_{code} \gets A_{ASA}Coding(Code_{Alg^*}, G')$
\STATE Get optimized solution $S^n$ : 
\STATE  \hspace{0.2cm}$S^0 \gets A_{ASA}(S_{code},Alg^*, G')$;
        \FOR {$i=1,2,\cdots,N_{check}$}
            \STATE  $S^n \gets A_{ASA}(S^{n-1},Alg^*, G')$;
        \ENDFOR
\end{algorithmic}
\end{algorithm}

\noindent\textbf{Graph Structure Information Extraction Agent (GSIEA)}: The GSIEA extracts implicitly embedded graph-structural information from text, particularly structured formats like tables, lists, adjacency matrices, or edge lists. It parses these inputs to identify nodes, edges, weights, and other topological properties, converting them into standardized graph representations (e.g., adjacency matrices, lists). This transformation enables downstream agents to efficiently use the extracted data for problem solving.

\noindent\textbf{Problem Information Extraction Agent (PIEA)}: The PIEA leverages LLMs’ problem classification capabilities to analyze real-world graph-theoretic problems, identify their types, and extract key components. It classifies problems (e.g., shortest path, network flow, graph matching), extracts relevant constraints and objectives, and outputs the information in a structured format. This guidance improves the accuracy and efficiency of downstream problem-solving agents. Formally,the operation of IEL is:
\begin{equation}
\small
\begin{gathered}
\mathcal{T} \gets A_{TIEA}(P), 
\mathcal{P} \gets A_{PIEA}(P), \\
G \gets  A_{GSIEA}(P),
IEL_{output} = (\mathcal{T},\mathcal{P},G)
\end{gathered}
\end{equation}
where $P$ is graph theory problem and $A_*$ is a different agent in IEL, $(\mathcal{T},\mathcal{P},G)$ represent the extracted text information, question information and graph structure information respectively.

\subsection{Knowledge Integration Layer (KIL)}
The primary objective of this layer is to construct structured graph data with high representational capacity and integrate graph-theoretic principles for advanced modeling, thus enhancing the efficiency of the solution and the quality of optimization.

\noindent\textbf{Structured Graph Information Agent (SGIA)}: The SGIA standardizes graph data from the GSIEA for efficient, consistent, and usable output. It cleans, deduplicates, and optimizes raw data into formats compatible with diverse environments to ensure accuracy. Additionally, it optimizes data storage and indexing based on algorithm requirements, enhancing computational efficiency for large-scale graphs. Without this agent, data inconsistencies, redundancy, and unoptimized structures could hinder algorithm performance. As a key component of MA-GTS, it ensures data standardization and optimization for efficient, scalable problem-solving.

\noindent\textbf{Graph Theory Agent (GTA)}: The GTA integrates information from the TIEA and PIEA with a Graph Theory Knowledge Base to analyze graph problems and find optimal solutions, improving LLM inference efficiency. It models the input problem by extracting key features such as type, constraints, and structural complexity, then queries the \textbf{Graph Theory Knowledge Base} to select the most suitable solution method from classical algorithms (e.g., shortest path, maximum flow, graph matching) \cite{gallo1988shortest,goldberg1988new} and heuristic techniques. By matching problems to algorithms, it reduces inefficient exhaustive searches, cutting computational costs and improving solution quality. Additionally, it also guides multi-agent collaboration, allowing the AEL to directly invoke optimal algorithms for efficient, scalable execution. Without it, LLMs risk poor strategy selection, high computation, and lower efficiency. As a key MA-GTS component, it ensures effective algorithm selection and inference in complex graph tasks. Formally,the operation of KIL is:
\begin{equation}
\small
\begin{gathered}
\mathcal{L}_{\mathcal{P}} \gets  A_{GTA}(\mathcal{T},\mathcal{P}, \mathcal{K_G}), \\
Alg^* \gets \arg\operatorname{opt}_{Alg_i \in \mathcal{L}_{\mathcal{P}}}  A_{GTA}(Alg_i, \mathcal{T}),\\
G' \gets A_{SGIA}(G),
KIL_{output} = (G', Alg^*)
\end{gathered}
\end{equation}
where $\mathcal{L}_{\mathcal{P}}$ represents the set of graph theory algorithms selected by GTA based on textual and problem-specific information, $\mathcal{K_G}$ denotes the Graph Theory Knowledge Base, $Alg^*$ refers to the algorithm suitable for the given graph size, and $G'$ stands for the normalized graph structure data.

\subsection{Algorithm Execution Layer (AEL)}
The primary goal of this layer is to integrate multiple algorithmic paradigms, ensuring efficient, scalable, and robust solutions under various constraints. Without it, the MA-GTS framework would rely solely on LLM-based inference, leading to high computational costs, instability, or suboptimal outcomes. As the computational core, the AEL enables the efficient solution of complex graph-theoretic problems across varying scales and complexities.

\noindent\textbf{Algorithm Solving Agent (ASA)}: The ASA is the core computational unit of the AEL, responsible for solving problems by executing algorithmic functions based on the optimal strategy selected by the GTA and the structured graph data processed by the SGIA. It utilizes a \textbf{Graph Theory Algorithm Library} that integrates exact algorithms \cite{noto2000method} and heuristic approaches, ensuring suitable solutions across various problem scenarios. After computation, the agent performs result integration and verification through cross-validation, error analysis, and constraint checking to ensure correctness. The ASA also offers explainable reasoning with inference paths, key decisions, and optimization steps for transparency. As MA-GTS's computational core, it delivers efficient, robust, and scalable solutions for complex graph problems. Formally,the operation of AEL is:
\begin{equation}
\small
\begin{gathered}
Code_{Alg^*} \gets A_{ASA}(Alg^*, \mathcal{L}_{code}), \\
S_{code} \gets A_{ASA}Coding(Code_{Alg^*}, G'),\\
S^0 \gets A_{ASA}(S_{code},Alg^*, G'),
\end{gathered}
\end{equation}
where $\mathcal{L}_{code}$ represents the Graph Theory Algorithm Library, $Code_{Alg^*}$ denotes the code obtained after optimal algorithm matching by ASA, $S_{code}$ refers to the output generated by running the code, and $S^0$ represents the interpretable output obtained by combining the code output with problem review. Finally, ASA undergoes $n$ rounds of self-checking, ultimately producing the final suitable result, $S^n$.

\section{G-REAL}
Existing datasets for evaluating LLMs' understanding and reasoning on graph-structured data are explicitly constructed. However, real-world graph-theoretic problems often involve rich textual semantic information and implicitly structured representations. To assess the performance of the MA-GTS framework on practical problems, we introduce \textbf{G-REAL}, a dataset that captures real-world graph problems. This dataset comprises four commonly encountered graph-theoretic challenges: the \textbf{Traveling Salesman Problem (TSP)}, the \textbf{Minimum Graph Coloring Problem}, the\textbf{ Minimum Vertex Cover Problem} and the \textbf{Shortest Path Problem}, respectively \cite{hoffman2013traveling,jensen2011graph,hochbaum1982approximation}. They correspond to four common problems in real-world scenarios, namely the \textbf{Delivery Logistics Problem}, the \textbf{Wireless Channel Allocation Problem}, the \textbf{Network Monitoring Problem}, and the \textbf{Target Navigation Problem}. The composition of G-REAL can be seen briefly in Figure \ref{fig:2}. In this section, we provide a detailed description of the dataset’s composition and construction methodology. More detail about G-REAL in Appendix \ref{sec:C}.

\begin{table}[t]
\centering
\begin{adjustbox}{max width=\linewidth} 
\begin{tabular}{c|cccc|c|cc}
\hline
\rowcolor[HTML]{EFEFEF} 
                                                    & \multicolumn{4}{c|}{\cellcolor[HTML]{EFEFEF}\textbf{G-REAL}} & \textbf{GraCoRe} & \multicolumn{2}{c}{\cellcolor[HTML]{EFEFEF}\textbf{NLGraph}} \\

\rowcolor[HTML]{EFEFEF} 
                                                     & \textbf{TSP} & \textbf{Coloring} & \textbf{Vetex Cover} & \textbf{Shortest Path} & \textbf{TSP}     & \textbf{Shortest Path} & \textbf{Cycle} \\ \hline
\cellcolor[HTML]{EFEFEF}\textbf{\#Graph}            & 900          & 900               & 900                  & 900                    & 360              & 380                    & 1150           \\
\cellcolor[HTML]{EFEFEF}\textbf{Node Range}        & 8 to 25      & 8 to 25           & 8 to 25              & 8 to 25                & 8 to 25          & 5 to 20                & 5 to 15        \\
\cellcolor[HTML]{EFEFEF}\textbf{Real-World Problem} & Delivery Logistics             & Wireless Channel Allocation       & Network Monitoring       & Target Navigation                & \XSolidBold                & \XSolidBold                                 & \XSolidBold                        \\
\cellcolor[HTML]{EFEFEF}\textbf{Text Noise}         & \CheckmarkBold              & \CheckmarkBold          & \CheckmarkBold          & \CheckmarkBold                     & \XSolidBold                & \XSolidBold                                 & \XSolidBold                        \\ \hline
\end{tabular}
\end{adjustbox}
     \caption{Differences between different datasets.}
    \label{tab:1}
         \vspace{-20pt}
\end{table}

\begin{table*}[t]
	\centering
	\scriptsize
	\setlength{\extrarowheight}{2.5pt} 
 \resizebox{\linewidth}{!}{
\begin{tabular}{c|c|cccc|c|cc}
\hline
\rowcolor[HTML]{EFEFEF} 
\cellcolor[HTML]{EFEFEF}                                            & \cellcolor[HTML]{EFEFEF}                                  & \multicolumn{4}{c|}{\cellcolor[HTML]{EFEFEF}\textbf{G-REAL}}                                        & \textbf{GraCoRe} & \multicolumn{2}{c}{\cellcolor[HTML]{EFEFEF}\textbf{NLGraph}} \\
\rowcolor[HTML]{EFEFEF} 
\multirow{-2}{*}{\cellcolor[HTML]{EFEFEF}\textbf{Model}}            & \multirow{-2}{*}{\cellcolor[HTML]{EFEFEF}\textbf{Method}} & \textbf{\makecell[c]{Delivery Logistics\\Problem (TSP)}}   & \textbf{\makecell[c]{Wireless Channel\\Allocation Problem (Coloring)}} & \textbf{\makecell[c]{Network Monitoring\\Problem (Vetex Cover)}} & \textbf{\makecell[c]{Target Navigation\\Problem (Shortest Path)}}                 & \textbf{TSP}     & \textbf{Shortest Path}           & \textbf{Cycle}            \\ \hline
\rowcolor[HTML]{FFFFFF} 
\cellcolor[HTML]{EFEFEF}                                            & \cellcolor[HTML]{EFEFEF}\textbf{Direct}                   & 11.8\%          & 80.1\%            & 68.7\%               & 47.1\%                                 & 79.7\%           & \textbf{100.0\%}                 & 97.3\%                    \\
\rowcolor[HTML]{FFFFFF} 
\multirow{-2}{*}{\cellcolor[HTML]{EFEFEF}\textit{o3-mini}}          & \cellcolor[HTML]{EFEFEF}\textbf{CoT}                      & 12.9\%          & 83.1\%            & 72.8\%               & 41.8\%                                 & 80.0\%           & 98.4\%                           & 97.8\%                    \\ \hline
\rowcolor[HTML]{FFFFFF} 
\cellcolor[HTML]{EFEFEF}                                            & \cellcolor[HTML]{EFEFEF}\textbf{Direct}                   & 2.5\%           & 23.4\%            & 0.3\%                & 7.1\%                                  & 1.1\%            & 27.3\%                           & 50.9\%                    \\
\rowcolor[HTML]{FFFFFF} 
\multirow{-2}{*}{\cellcolor[HTML]{EFEFEF}\textit{GPT-4o-mini}}      & \cellcolor[HTML]{EFEFEF}\textbf{CoT}                      & 3.1\%           & 25.1\%            & 0.0\%                & 6.4\%                                  & 1.1\%            & 27.6\%                           & 51.1\%                    \\ \hline
\rowcolor[HTML]{FFFFFF} 
\cellcolor[HTML]{EFEFEF}                                            & \cellcolor[HTML]{EFEFEF}\textbf{Direct}                   & 0.1\%           & 0.7\%             & 2.5\%                & 4.0\%                                  & 1.9\%            & 30.5\%                           & 50.0\%                    \\
\rowcolor[HTML]{FFFFFF} 
\multirow{-2}{*}{\cellcolor[HTML]{EFEFEF}\textit{GPT-3.5}}          & \cellcolor[HTML]{EFEFEF}\textbf{CoT}                      & 2.1\%           & 7.6\%             & 4.8\%                & 3.6\%                                  & 1.6\%            & 34.7\%                           & 49.9\%                    \\ \hline
\rowcolor[HTML]{FFFFFF} 
\cellcolor[HTML]{EFEFEF}                                            & \cellcolor[HTML]{EFEFEF}\textbf{Direct}                   & 0.6\%           & 16.2\%            & 17.4\%               & 4.6\%                                  & 3.8\%            & 22.1\%                           & 49.6\%                    \\
\rowcolor[HTML]{FFFFFF} 
\multirow{-2}{*}{\cellcolor[HTML]{EFEFEF}\textit{Qwen2.5-7B}}       & \cellcolor[HTML]{EFEFEF}\textbf{CoT}                      & 0.6\%           & 8.8\%             & 8.5\%                & 5.8\%                                  & 3.0\%            & 27.3\%                           & 52.7\%                    \\ \hline
\rowcolor[HTML]{FFFFFF} 
\cellcolor[HTML]{EFEFEF}                                            & \cellcolor[HTML]{EFEFEF}\textbf{Direct}                   & 3.6\%           & 10.1\%            & 7.2\%                & 4.3\%                                  & 0.3\%            & 12.6\%                           & 53.7\%                    \\
\rowcolor[HTML]{FFFFFF} 
\multirow{-2}{*}{\cellcolor[HTML]{EFEFEF}\textit{Llama3-7B}}        & \cellcolor[HTML]{EFEFEF}\textbf{CoT}                      & 4.1\%           & 14.3\%            & 6.7\%                & 4.2\%                                  & 0.3\%            & 19.4\%                           & 50.9\%                    \\ \hline
\rowcolor[HTML]{FFFFFF} 
\cellcolor[HTML]{EFEFEF}                                            & \cellcolor[HTML]{EFEFEF}\textbf{Direct}                   & 4.9\%           & 27.2\%            & 21.1\%               & 11.4\%                                 & 10.5\%           & 50.8\%                           & 78.1\%                    \\
\rowcolor[HTML]{FFFFFF} 
\multirow{-2}{*}{\cellcolor[HTML]{EFEFEF}\textit{Deepseek-V3-660B}} & \cellcolor[HTML]{EFEFEF}\textbf{CoT}                      & 5.5\%           & 28.3\%            & 22.2\%               & 32.2\%                                 & 18.8\%           & 92.9\%                           & 77.8\%                    \\ \hline
\rowcolor[HTML]{FFFFFF} 
\cellcolor[HTML]{EFEFEF}\textit{OWL (GPT-4o-mini)}                   & \cellcolor[HTML]{EFEFEF}\textbf{Multi-Agent}              & 10.2\%          & 47.4\%            & 7.8\%                & 19.1\%                                 & 4.4\%            & 36.3\%                           & 49.7\%                    \\ \hline
\rowcolor[HTML]{FFFFFF} 
\cellcolor[HTML]{EFEFEF}\textit{GraphTeam (GPT-4o-mini)}             & \cellcolor[HTML]{EFEFEF}\textbf{Multi-Agent}              & 8.8\%           & 90.0\%            & 12.0\%               & 87.7\%                                 & 84.4\%           & 98.4\%                           & {\color[HTML]{000000} \textbf{100.0\%}  }   \\ \hline
\rowcolor[HTML]{FFFFFF} 
\cellcolor[HTML]{EFEFEF}\textit{\textbf{MA-GTS (Deepseek-V3-660B)}}       & \cellcolor[HTML]{EFEFEF}\textbf{Multi-Agent}              & 76.2\%          & 88.2\%            & {\color[HTML]{000000} \textbf{99.1\%(\color[HTML]{CB0000}$\uparrow$26.3\%)}}      & 88.2\%                                 & 93.1\%           & 93.8\%                           & {\color[HTML]{000000} \textbf{100.0\%} }         \\ \hline
\rowcolor[HTML]{FFFFFF} 
\cellcolor[HTML]{EFEFEF}\textit{\textbf{MA-GTS (GPT-4o-mini)}}       & \cellcolor[HTML]{EFEFEF}\textbf{Multi-Agent}              & {\color[HTML]{000000} \textbf{94.9\%(\color[HTML]{CB0000}$\uparrow$82\%)}}              & {\color[HTML]{000000} \textbf{94.5\%(\color[HTML]{CB0000}$\uparrow$4.5\%)}}                 & 93.2\%           & {\color[HTML]{000000} \textbf{91.7\%(\color[HTML]{CB0000}$\uparrow$4\%)}} & {\color[HTML]{000000} \textbf{96.9\%(\color[HTML]{CB0000}$\uparrow$12.5\%)}} & 97.8\%(\color[HTML]{34FF34}$\downarrow$2.2\%)                          & 98.9\%                    \\ \hline
\end{tabular}

}
\vspace{-5pt}
     \caption{The performance comparison of LLMs and MA-GTS on G-REAL and two benchmarks is shown. Red text indicates MA-GTS’s accuracy improvement over the best LLM, while green text highlights the opposite. GPT-4o-mini was used as the base model for MA-GTS.}
    \label{tab:2}
         \vspace{-15pt}
\end{table*}
\subsection{Data Collection}
To mitigate the risk of data contamination in LLMs, which could lead to biased test accuracy due to prior exposure to training data, G-REAL employs several techniques, including randomized node naming, synthetic node descriptions, added textual noise, and randomly structured graph representations. Node names are generated by randomly combining the 26 letters of the alphabet, and synthetic node descriptions are created with arbitrary textual representations. For example, a node may be described as: \textit{"Amber Plaza: A bustling central square surrounded by cafes, boutiques, and street performers."} These fictional descriptions ensure that LLMs cannot leverage prior knowledge, maintaining the integrity of the evaluation.

To improve dataset realism and obscure graph structure, we introduce textual noise to each instance, simulating real-world graph problems embedded in unstructured text. Graph structures are randomly generated, with each node assigned a unique name to reduce prior LLM exposure. Optimal and approximate solutions are generated for each problem type using established algorithms, providing benchmarks for evaluating both LLM and MA-GTS performance.

\subsection{Data Statistics }
To evaluate our framework’s effectiveness in real-world graph-theoretic problems, we construct test datasets with graph sizes from 8 to 25 nodes for each problem type. Each sub-dataset includes 50 instances with distinct structures, offering both optimal and approximate solutions for a comprehensive assessment of robustness and generalization. A statistical summary is provided in Table \ref{tab:1}.

\subsection{Evaluation}
For the TSP, Minimum Graph Coloring, Minimum Vertex Cover, and Shortest Path problems, the output includes both selected nodes and the final solution, requiring dual evaluation. To fully assess LLMs’ graph reasoning, both output types are used as evaluation metrics. The model's performance is measured by verifying the accuracy of both the selected node set and the computed solution. The methodology for calculating the final accuracy is as follows:
$\text{$ACC_{ALL}$} = 0.5 \cdot ACC_{\text{nodes}} + 0.5 \cdot ACC_{\text{result}},$ where $ACC_{\text{nodes}}$ and $ACC_{\text{result}}$ represent the accuracy of the node set and the predicted values, respectively, with a value of 1 for correct predictions and 0 for incorrect ones.


\section{Experiments Setup}
\subsection{Datasets}
To evaluate the reasoning capabilities of the MA-GTS framework across various graph-theoretic problem types, complexities, and domains, we used the \textbf{G-REAL} dataset alongside two benchmark datasets, \textbf{GraCoRe} \cite{yuan2025gracore} and \textbf{NLGraph} \cite{NLgraph-6}, covering seven distinct graph-theoretic tasks. We selected three sub-tasks for evaluation: the TSP, shortest path problem, and Cycle problem in GraCoRe and NLGraph. Notably, both GraCoRe and G-REAL include TSP instances, both NLGraph and G-REAL include Shortest Path instances; however, the G-REAL TSP and Shortest Path is more complex and reflects real-world scenarios with implicit graph structure data. By comparing performance on these two types instances, we assess the model's ability to handle more intricate problems. The simpler tasks in NLGraph evaluate the generalization and robustness of MA-GTS. A summary of the differences between these datasets is provided in Table \ref{tab:1}.
\subsection{Baselines and Foundation Model}
We compared three of OpenAI's latest closed-source models:  \textit{o3-mini},  \textit{GPT-4o-mini}, and  \textit{GPT-3.5} \cite{gpt4}. Additionally, we evaluated three of the most recent open-source models:  \textit{Llama3-7b} \cite{llama}, \textit{Qwen2.5-7b} \cite{qwen} and Deepseek-V3-660B \cite{liu2024deepseek}. For the evaluation methodology, we adopted both direct inference and CoT reasoning approaches. For the foundation model, we selected the GPT-4o-mini and Deepseek-V3-660B model, they respectively represent some of the more advanced open-source and closed-source models. Furthermore, we conducted a comparative analysis of the performance of OWL \cite{owl2025} and GraphTeam \cite{li2024graphteam}, which respectively represent a general-purpose multi-agent framework and a graph-theoretic multi-agent framework. Regarding the final test results, for each task, we used the accuracy of the final computed solution as the primary evaluation metric. More details about models in Appendix \ref{sec:B}.
\section{Results and Analysis}
We evaluate the performance of our framework against other LLMs on graph theory problems, with results presented in Table \ref{tab:2}. MA-GTS outperforms all baselines, achieving state-of-the-art results and matching the performance of the leading o3-mini model on simpler problems. We also assess the MA-GTS framework from multiple perspectives.
\begin{table*}[t]
	\centering
	\scriptsize
	\setlength{\extrarowheight}{2.5pt} 
     \resizebox{\linewidth}{!}{
\begin{tabular}{c|ccc|ccc|ccc|ccc}
\hline
\rowcolor[HTML]{EFEFEF} 
                                                              & \multicolumn{3}{c|}{\cellcolor[HTML]{EFEFEF}\textbf{TSP}}                 & \multicolumn{3}{c|}{\cellcolor[HTML]{EFEFEF}\textbf{Coloring}}            & \multicolumn{3}{c|}{\cellcolor[HTML]{EFEFEF}\textbf{VertexCover}}          & \multicolumn{3}{c}{\cellcolor[HTML]{EFEFEF}\textbf{Shortest Path}}\\
\rowcolor[HTML]{EFEFEF} 
                                                              & \textbf{Inp.Tokens(k)} & \textbf{Out.Tokens(k)} & \textbf{Price(\$)}      & \textbf{Inp.Tokens(k)} & \textbf{Out.Tokens(k)} & \textbf{Price(\$)}          & \textbf{Inp.Tokens(k)} & \textbf{Out.Tokens(k)} & \textbf{Price(\$)}   
                                                              & \textbf{Inp.Tokens(k)} & \textbf{Out.Tokens(k)} & \textbf{Price(\$)} \\ \hline
\cellcolor[HTML]{EFEFEF}\textit{o3-mini}                      & 2.31                   & 4.98                   & 0.0244                  & 0.69                   & 6.83                   & 0.0309                  & 0.6                    & 9.73                   & 0.0443    & 0.87                    & 3.12                   & 0.0147                \\
\cellcolor[HTML]{EFEFEF}\textit{\textbf{MA-GTS(GPT-4o-mini)}} & 13.32                  & \textbf{4.56\color[HTML]{CB0000}($\downarrow$8.4\%)}   & \textbf{0.0047\color[HTML]{CB0000}($\downarrow$80.7\%)} & 6.79                   & \textbf{2.57\color[HTML]{CB0000}($\downarrow$62.4\%)}  & \textbf{0.0025\color[HTML]{CB0000}($\downarrow$91.9\%)} & 6.39                   & \textbf{2.31\color[HTML]{CB0000}($\downarrow$76.2\%)}  & \textbf{0.0023\color[HTML]{CB0000}($\downarrow$94.8\%)} & 7.36 & \textbf{2.42\color[HTML]{CB0000}($\downarrow$22.4\%)}& \textbf{0.0025\color[HTML]{CB0000}($\downarrow$82.9\%)}\\ \hline
\end{tabular}}
\vspace{-10pt}
     \caption{Comparison of inference costs between MA-GTS and o3-mini model on G-REAL.}
    \label{tab:3}
      \vspace{-15pt}
\end{table*}
\subsection{Performance on real-world problems}
As shown in Table \ref{tab:2}, G-REAL provides four real-world graph theory problems, with TSP being the most complex. Based on the results from these problems, MA-GTS demonstrates superior performance, achieving an accuracy rate exceeding 90\% across all tests. Notably, in the case of the TSP, MA-GTS outperforms the o3-mini model by 82\%. 
Even when built upon the open-source DeepSeek model, MA-GTS still achieves strong performance. Furthermore, when compared to the GPT-4o-mini model, MA-GTS significantly improves its performance from 3.1\% to 94.9\%, marking a substantial increase. This clearly underscores the effectiveness of our framework. Additionally, it is evident that, aside from the o3-mini model, other models exhibit subpar performance on the G-REAL dataset. It is particularly interesting that the performance gap between the two open-source and two closed-source models is minimal, suggesting that the complexity of the problems may lead to a consistent decline in performance, an issue that warrants further investigation. Overall, MA-GTS stands out for its advanced capabilities and generalization when handling complex graph theory problems.
\subsection{Performance on simple problem}
Table \ref{tab:2} shows that for simpler graph theory problems, such as the Shortest Path and Cycle problems from the NLGraph dataset, the o3-mini model performs exceptionally well, with MA-GTS also showing strong results. Specifically, for Shortest Path problem, the gap between MA-GTS and o3-mini is just 2.2\%, and MA-GTS performs equally well on the Cycle problem. In contrast, other models perform less satisfactorily. The MA-GTS framework, based on the GPT-4o-mini model, significantly enhances the accuracy of the 4o model, bringing it on par with the o3-mini. Overall, MA-GTS demonstrates excellent performance across diverse textual descriptions and graph structures, highlighting its remarkable generalization capabilities.
\subsection{G-REAL effectiveness analysis}
To evaluate the performance of LLMs and MA-GTS on real-world graph theory problems, we constructed the G-REAL dataset. As shown in Table \ref{tab:2}, the performance of existing LLMs on the G-REAL dataset is suboptimal. To validate the effectiveness of this dataset, we compared it with the TSP problem from the GraCoRe Benchmark, testing problems with node sizes ranging from 8 to 25, consistent with the scale of G-REAL. From this comparison, we observe that on the G-REAL dataset, which includes text complexity, added text noise, and node name shuffling, the o3-mini model performs poorly, with its accuracy dropping from 79.7\% in GraCoRe to 11.8\%. In contrast, the MA-GTS framework appears unaffected by the complexities of real-world graph theory problems, maintaining performance above 90\%. This result indirectly supports the validity of the G-REAL dataset and shows the stability of the MA-GTS.
\begin{figure}[t]
\vspace{-5pt}
\centering
  \includegraphics[width=0.5\textwidth]{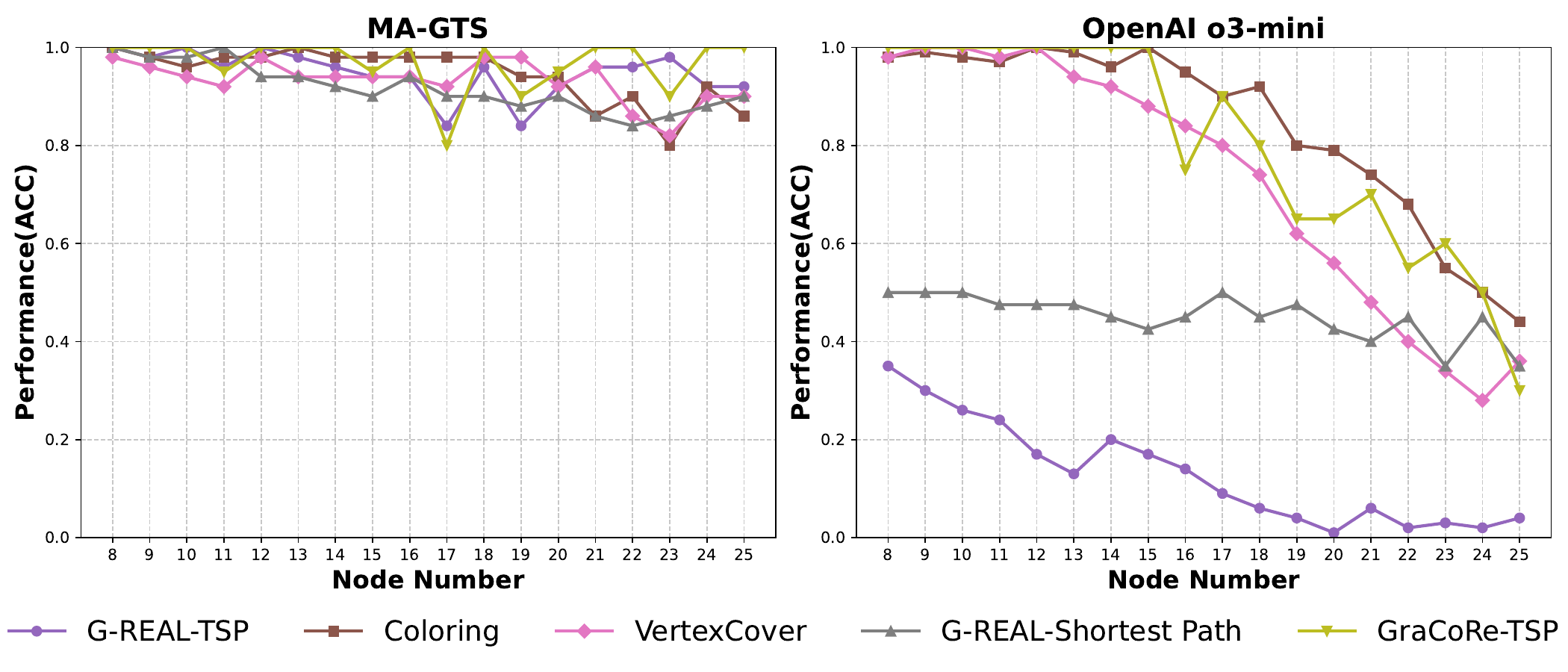}
  \vspace{-20pt}
  \caption {Performance of different problems across varying node numbers (MA-GTS v.s. o3-mini).}
    \label{fig:3} 
          \vspace{-15pt}
\end{figure}
\vspace{-10pt}
\subsection{Impact of Node Size}
To evaluate the impact of node scale on LLMs in complex graph theory problems, we tested the performance of MA-GTS and the o3-mini model on four complex graph problem datasets, with node sizes ranging from 8 to 25. The results, shown in Figure \ref{fig:3}, clearly demonstrate that as the number of nodes increases, the performance of the o3-mini model deteriorates, particularly in the TSP problem from G-REAL. For node sizes greater than 20, the o3-mini model is unable to produce correct answers. In contrast, under the MA-GTS, the effect of node size is less pronounced. Even with more than 20 nodes, MA-GTS maintains high prediction accuracy and stability. It highlights both the effectiveness and superiority of MA-GTS. Performance of MA-GTS on larger node scales is discussed in the Appendix \ref{sec:D}.
\subsection{Cost Analysis}
Since MA-GTS requires multiple agent calls to model APIs for inference, cost considerations arise. To address this, we compared the inference costs of MA-GTS based on the GPT-4o-mini model with the o3-mini model, as shown in Table \ref{tab:3}. Surprisingly, MA-GTS incurs significantly lower costs than the o3-mini model. The o3-mini model, in contrast, has hidden reasoning tokens during inference, leading to long, concealed reasoning processes even in direct inference scenarios. As shown in the table, the inference cost of MA-GTS is about one-tenth to one-twentieth of the o3-mini model, requiring far fewer inference tokens. Moreover, MA-GTS achieves far better results than o3-mini, demonstrating its high cost-effectiveness in delivering more accurate outcomes at a lower cost. Runtime efficiency is discussed in the Appendix \ref{sec:D}.
\begin{table}[t]
\centering
\begin{adjustbox}{max width=\linewidth} 
\begin{tabular}{
>{\columncolor[HTML]{EFEFEF}}c |
>{\columncolor[HTML]{FFFFFF}}c 
>{\columncolor[HTML]{FFFFFF}}c 
>{\columncolor[HTML]{FFFFFF}}c |
>{\columncolor[HTML]{FFFFFF}}c }
\hline
                               & \multicolumn{3}{c|}{\cellcolor[HTML]{EFEFEF}\textbf{G-REAL}}                                                                    & \cellcolor[HTML]{EFEFEF}                                              \\
                               & \cellcolor[HTML]{EFEFEF}\textbf{TSP} & \cellcolor[HTML]{EFEFEF}\textbf{Coloring} & \cellcolor[HTML]{EFEFEF}\textbf{Vetex Cover} & \multirow{-2}{*}{\cellcolor[HTML]{EFEFEF}\textbf{Average error rate}} \\ \hline
\textit{GPT-4o-mini(Tool use)} & 30.8\%                               & 39.0\%                                    & 4.6\%                                        & 75.0\%                                                                \\ \hline
\textit{w/o IEL}               & 12.5\%                               & 42.2\%                                    & 14.6\%                                       & 19.4\%                                                                \\
\textit{w/o KIL}               & 7.8\%                                & 37.1\%                                    & 12.8\%                                       & 1.0\%                                                                 \\
\textit{w/o AEL}               & 4.6\%                                & 32.1\%                                    & 7.4\%                                        & 3.2\%                                                                 \\ \hline
\textbf{MA-GTS(GPT-4o-mini)}   & \textbf{94.9\%}                      & \textbf{94.5\%}                           & \textbf{93.2\%}                              & \textbf{0.5\%}                                                        \\ \hline
\end{tabular}
\end{adjustbox}
     \caption{Ablation Experiments for Each Layer of MA-GTS ("Tool use" refers to the utilization of only the algorithm library we have constructed).}
    \label{tab:4}
\end{table}
\subsection{Ablations Studies and Analyses}
To validate the effectiveness of each layer in MA-GTS, we conducted ablation experiments, with results shown in Table \ref{tab:4}. It demonstrates that each layer is crucial, and removing any layer significantly affects the final results. Although the IEL layer has the smallest impact on accuracy, its absence leads to a substantial increase in error rate (19\%), highlighting its role in maintaining stability. The absence of the AEL layer results in the greatest accuracy loss. Even when a module is removed, MA-GTS still improves the accuracy of the base model, validating the framework's effectiveness. Additionally, when inference is performed using only the GPT-4o-mini model with the constructed algorithm library, accuracy improves, but the error rate remains high (75\%). For graph sizes larger than 10 nodes, the model struggles to correctly invoke algorithms, further demonstrating the robustness and generalizability of MA-GTS.
\begin{table}[t]
\centering
\begin{adjustbox}{max width=\linewidth} 
\begin{tabular}{c|ccccc|c|c|c}
\hline
\rowcolor[HTML]{EFEFEF} 
{\color[HTML]{333333} \textbf{}}                                                                                               & {\color[HTML]{333333} \textbf{Time 1}} & {\color[HTML]{333333} \textbf{Time 2}} & {\color[HTML]{333333} \textbf{Time 3}} & {\color[HTML]{333333} \textbf{Time 4}} & {\color[HTML]{333333} \textbf{Time 5}} & {\color[HTML]{333333} \textbf{Var}} & {\color[HTML]{333333} \textbf{Std}} & {\color[HTML]{333333} \textbf{Mean}} \\ \hline
\cellcolor[HTML]{EFEFEF}{\color[HTML]{2C3A4A} \textbf{\begin{tabular}[c]{@{}c@{}}Delivery \\ Logistics\end{tabular}}}          & 100\%                                  & 100\%                                  & 98.1\%                                 & 98.1\%                                 & 96.3\%                                 & 2.4                                 & 1.5                                 & 98.5\%                               \\ \hline
\cellcolor[HTML]{EFEFEF}{\color[HTML]{2C3A4A} \textbf{\begin{tabular}[c]{@{}c@{}}Wireless Channel \\ Allocation\end{tabular}}} & 100\%                                  & 98.1\%                                 & 94.4\%                                 & 96.3\%                                 & 92.4\%                                 & 8.9                                 & 2.9                                 & 96.2\%                               \\ \hline
\cellcolor[HTML]{EFEFEF}{\color[HTML]{2C3A4A} \textbf{\begin{tabular}[c]{@{}c@{}}Network \\ Monitoring\end{tabular}}}          & 100\%                                  & 100\%                                  & 100\%                                  & 100\%                                  & 100\%                                  & 0                                   & 0                                   & 100\%                                \\ \hline
\cellcolor[HTML]{EFEFEF}{\color[HTML]{2C3A4A} \textbf{\begin{tabular}[c]{@{}c@{}}Target \\ Navigation\end{tabular}}}           & 92.4\%                                 & 96.3\%                                 & 90.6\%                                 & 92.4\%                                 & 92.4\%                                 & 4.4                                 & 2.1                                 & 92.8\%                               \\ \hline
\end{tabular}
\end{adjustbox}
     \caption{Sensitivity analysis of multi-round inference.}
    \label{tab:5}
  \vspace{-15pt}
\end{table}
\subsection{Sensitivity Studies of MA-GTS}
We have conducted an additional sensitivity experiment for MA-GTS. For each graph dataset size, we randomly selected 5 problems and queried MA-GTS 5 times per problem to test the stability and sensitivity of the framework. As shown in Table \ref{tab:5}, the results demonstrate that MA-GTS is highly stable and there is no significant performance fluctuation across repeated queries.

\section{Conclusion}
We introduces MA-GTS, a Multi-Agent Framework for solving real-world graph theory problems, validated using the G-REAL dataset. Performance comparisons across various LLMs show that MA-GTS achieves high accuracy, stability, and cost-effectiveness, excelling in both complex and simpler graph problems. With accuracy consistently above 90\%, MA-GTS outperforms existing methods, maintaining stability across different problem scales and being well-suited for larger graphs. Future work will focus on scaling to even larger problems and improving cost-efficiency.
\section*{Limitations}
Although the MA-GTS framework demonstrates significant advantages in addressing complex graph-theoretic problems, several limitations remain. First, while the G-REAL dataset provides valuable support for validating the framework's effectiveness, it may not fully capture the diversity of real-world graph problems, thus limiting the generalizability of the framework. Second, the MA-GTS framework may still require substantial computational resources when handling large-scale problems, particularly in resource-constrained environments. Moreover, despite the improvements made in enhancing LLMs' graph structure modeling capabilities, LLMs may still encounter performance bottlenecks when dealing with graphs that exhibit highly dependent relationships or specialized structures. Finally, the current capabilities of open-source model invocation tools are insufficient, which may impact the stability of the MA-GTS framework.
\section*{Acknowledgments}
This work is supported by the National Science Foundation of China (U22B2059, 62276083), the Major Key Project of PCL (Grant No. PCL2024A08) and the 5G Application Innovation Joint Research Institute's Project (A003).

\bibliography{custom}
\clearpage
\appendix

\section{MA-GTS Details}
\label{sec:A}
\subsection{Graph Theory Knowledge Base}
\label{subsec:A1}
\begin{figure*}
\centering
  \includegraphics[width=\textwidth]{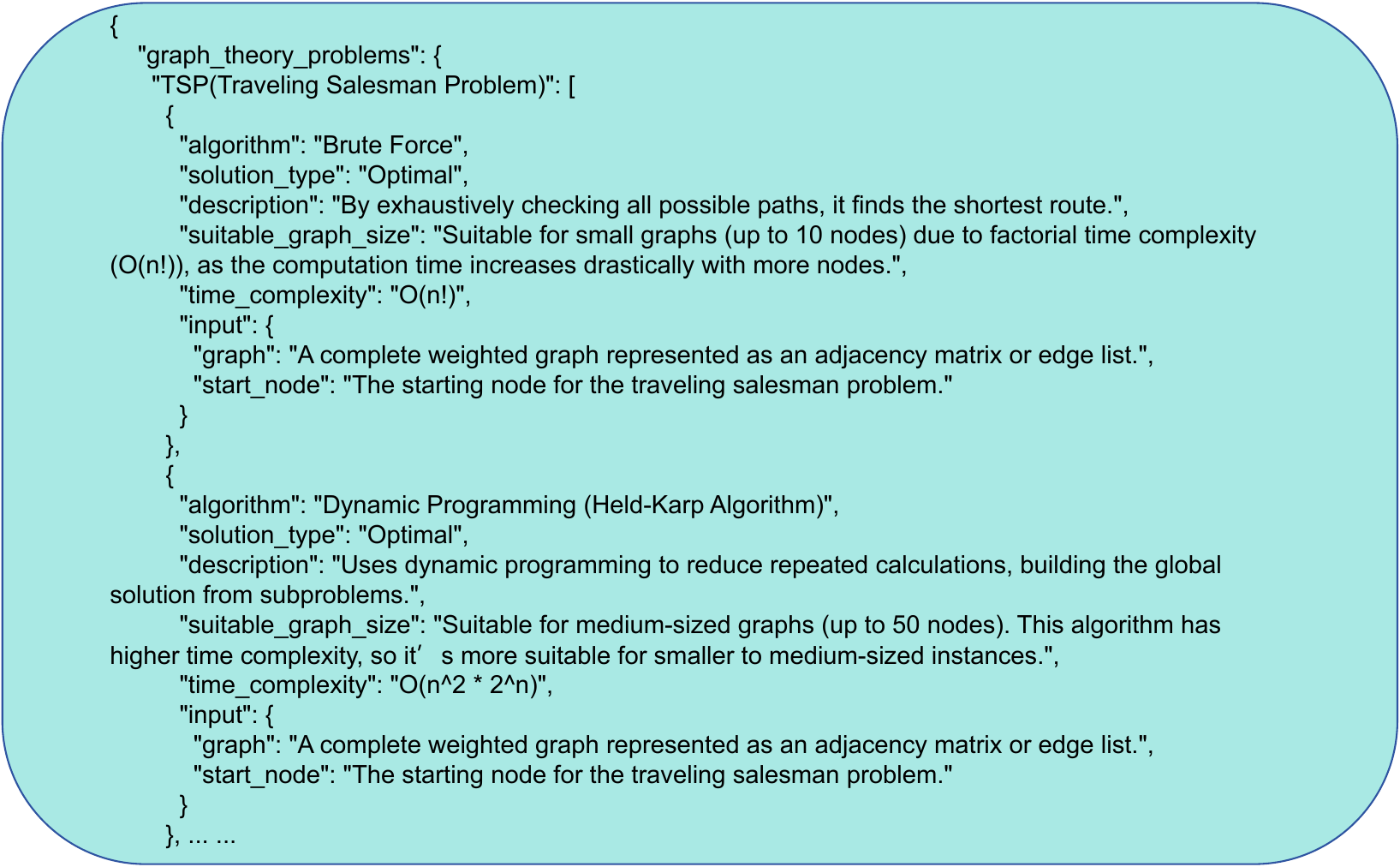}
  \caption {Details of Graph Theory Knowledge Base}
    \label{fig:4} 
\end{figure*}

The Graph Theory Knowledge Base is a graph theory problem database that we have constructed, containing a wide range of common graph theory problems encountered in daily life, including both complex and simple ones. Each problem is associated with multiple optimal or approximate solution algorithms. For each algorithm, we provide a detailed description of its complexity, applicable conditions, and parameter settings, though it does not include corresponding code. This database can serve as a reference book for agents in graph theory. A specific example can be seen in Figure \ref{fig:4}.
\subsection{Graph Theory Algorithm Library}
\label{subsec:A2}
\begin{figure*}
\centering
  \includegraphics[width=\textwidth]{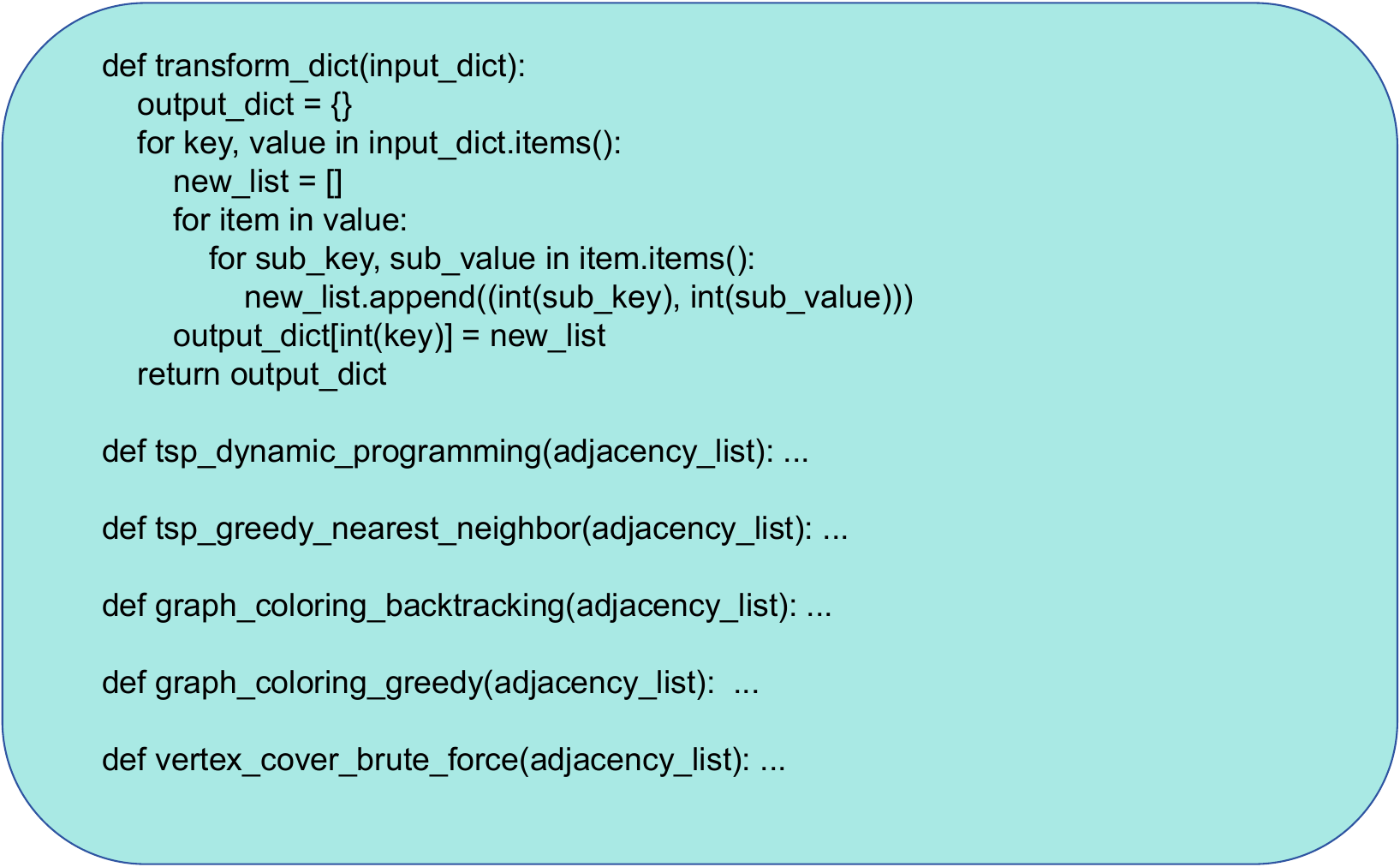}
  \caption {Details of Graph Theory Algorithm Library}
    \label{fig:5} 
\end{figure*}

The Graph Theory Algorithm Library is a Python code repository that we have constructed, containing code corresponding to the graph theory algorithms in the Graph Theory Knowledge Base. This ensures the correctness of input parameters and helps maintain the stability of the MA-GTS framework. Each code snippet is accompanied by detailed parameter descriptions and is designed to accommodate various types of graph structure representations. A specific example can be seen in Figure \ref{fig:5}.
\subsection{Prompt Templates}
\begin{figure*}
\centering
  \includegraphics[width=\textwidth]{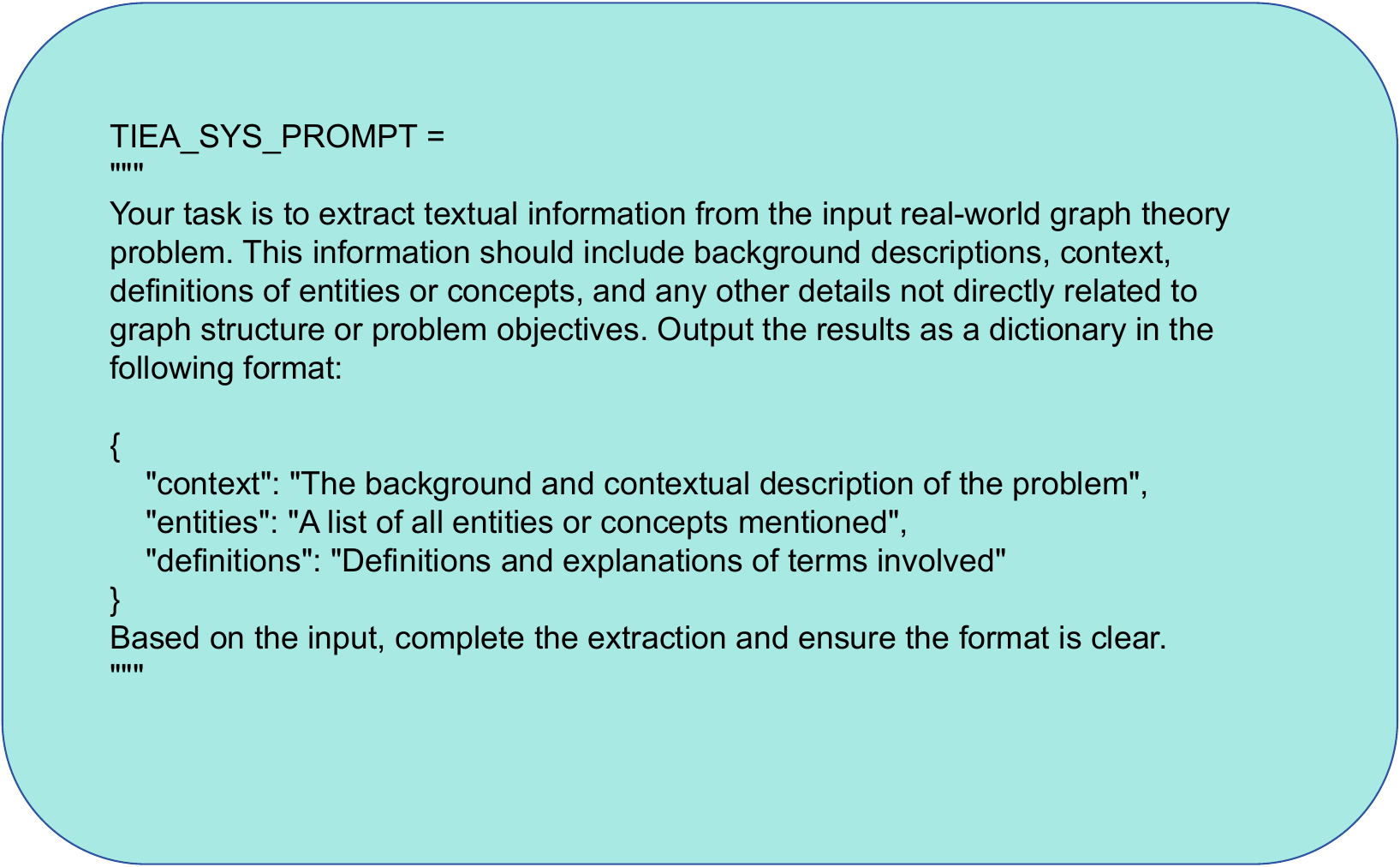}
  \caption {Details of TIEA}
    \label{fig:6} 
\end{figure*}

\begin{figure*}
\centering
  \includegraphics[width=\textwidth]{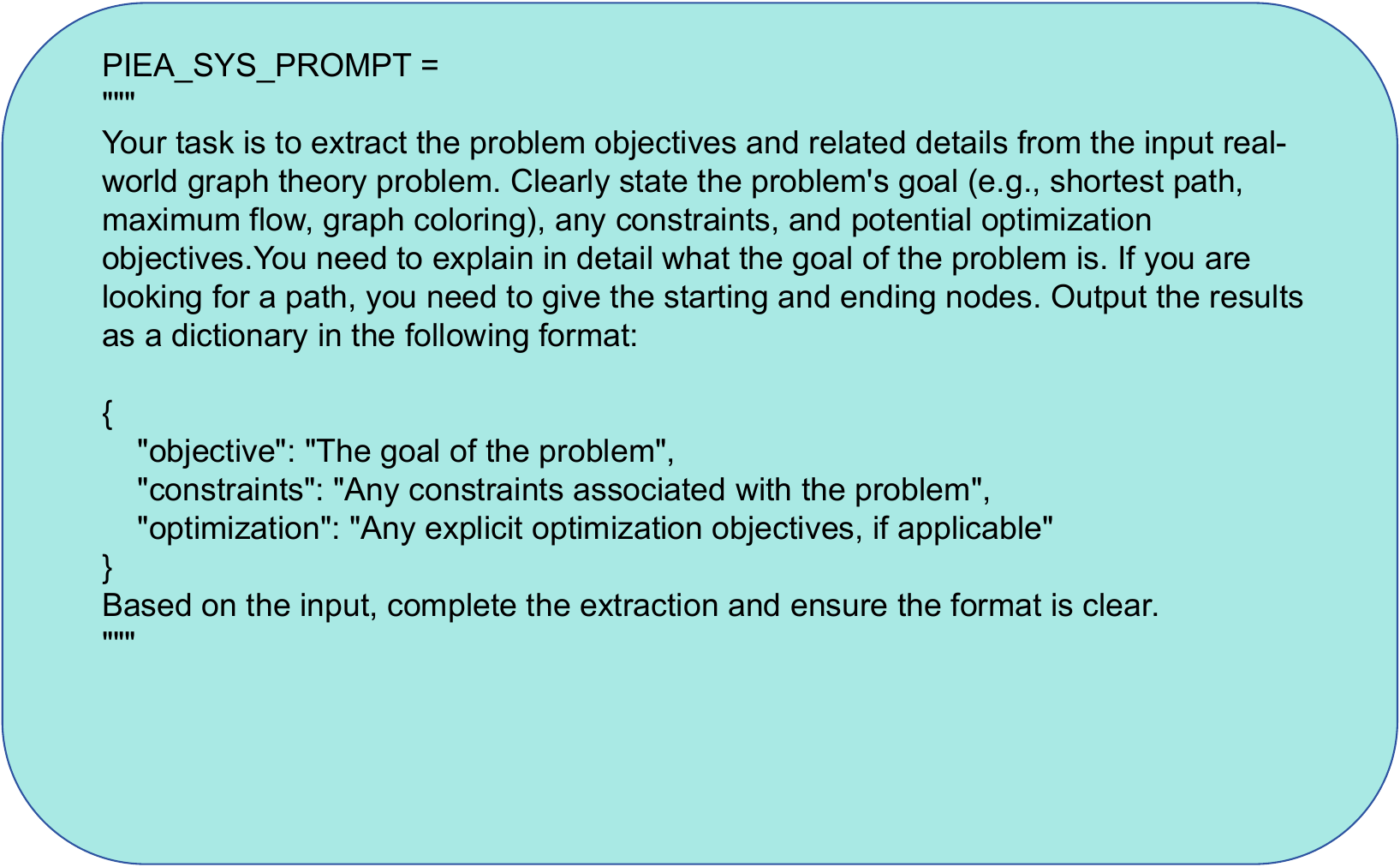}
  \caption {Details of PIEA}
    \label{fig:7} 
\end{figure*}

\begin{figure*}
\centering
  \includegraphics[width=\textwidth]{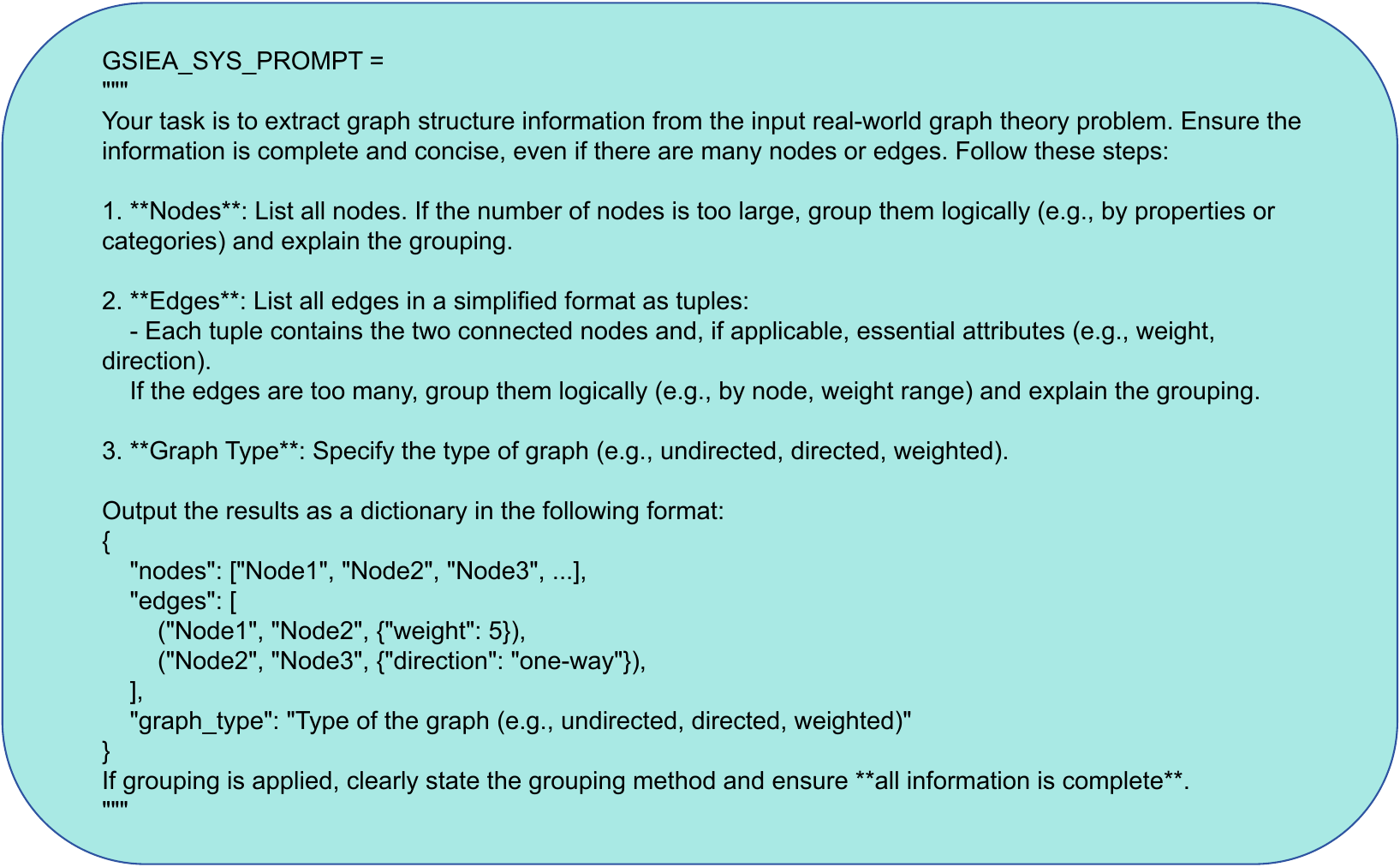}
  \caption {Details of GSIEA}
    \label{fig:8} 
\end{figure*}

\begin{figure*}
\centering
  \includegraphics[width=\textwidth]{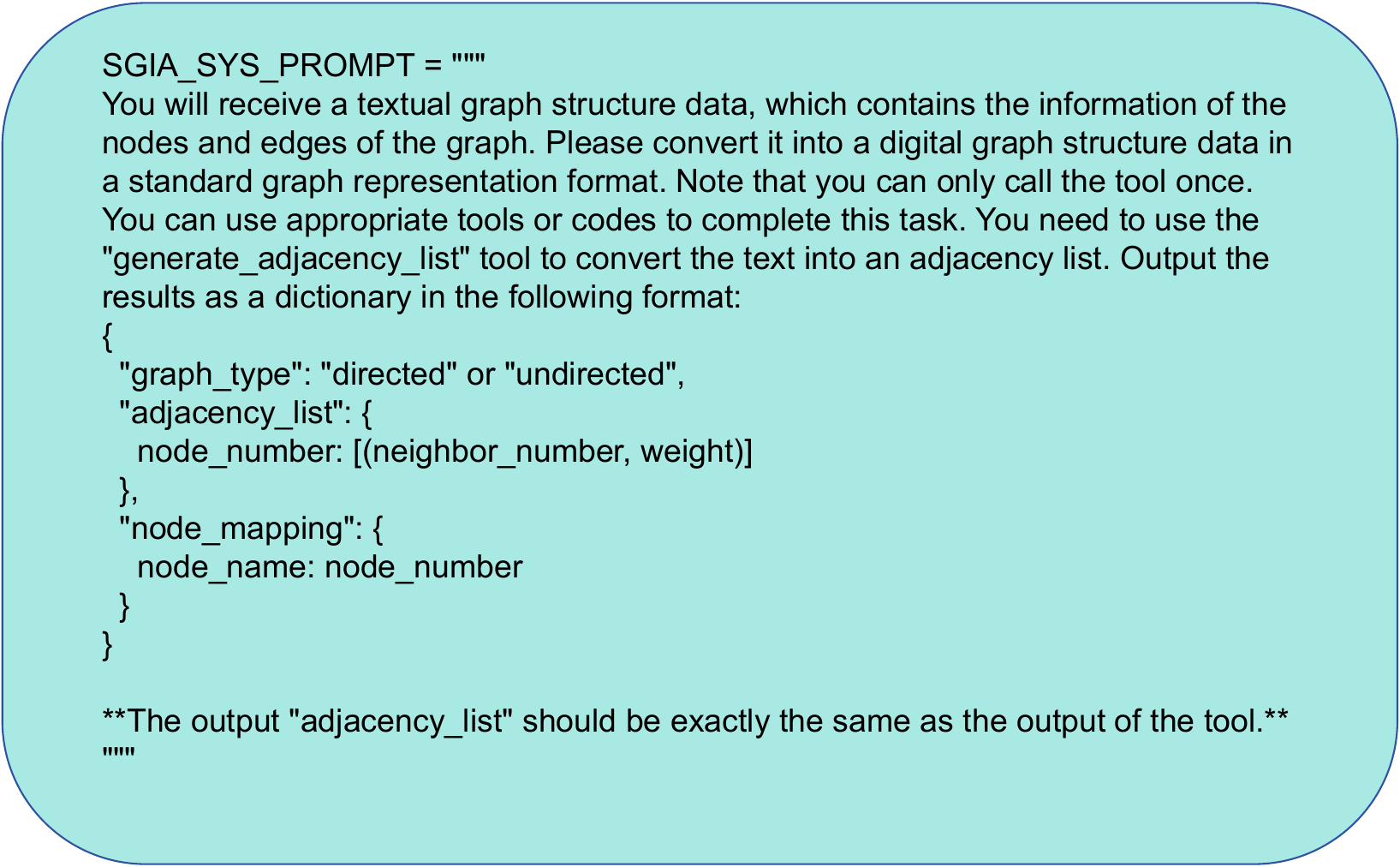}
  \caption {Details of SGIA}
    \label{fig:9} 
\end{figure*}

\begin{figure*}
\centering
  \includegraphics[width=\textwidth]{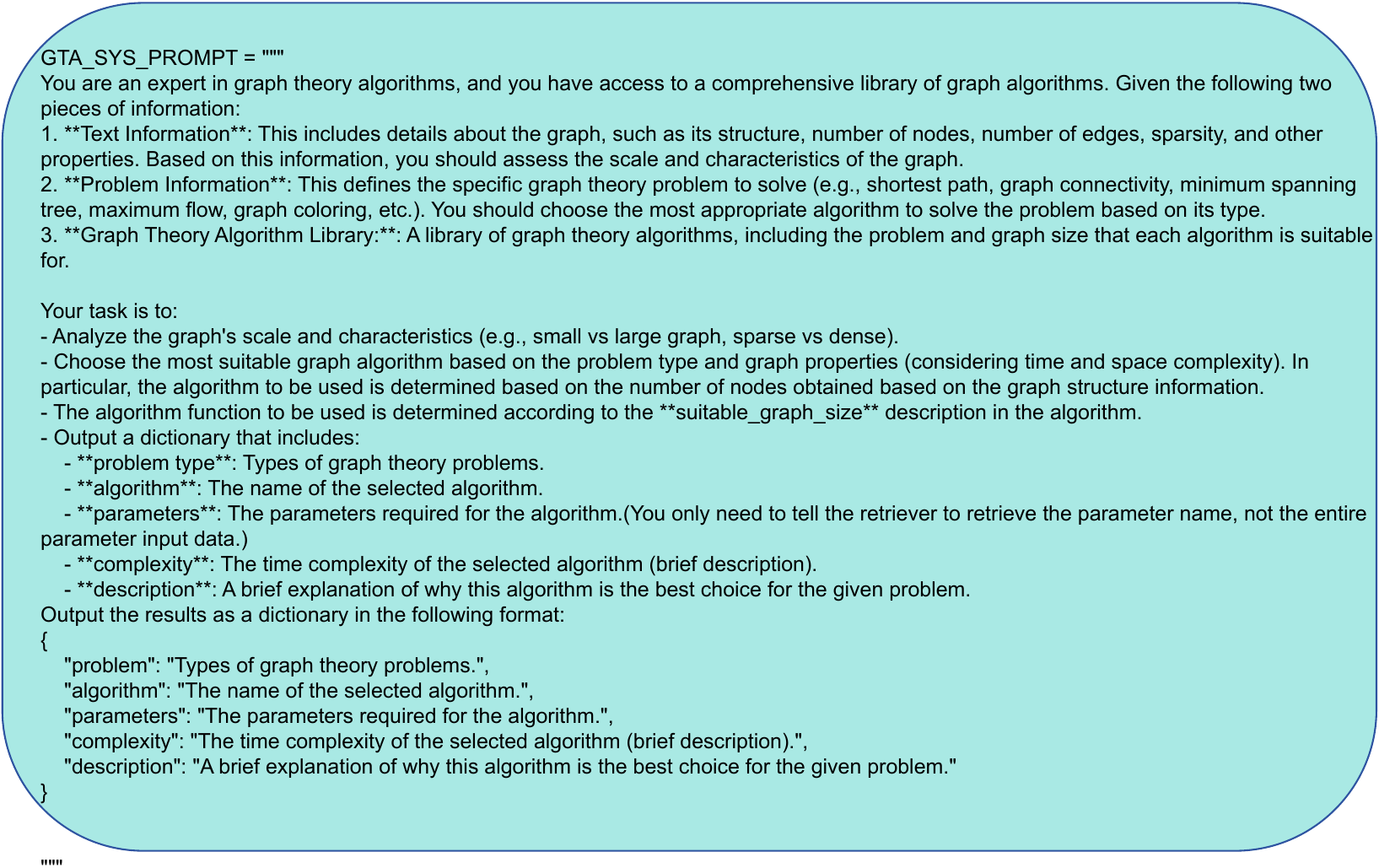}
  \caption {Details of GTA}
    \label{fig:10} 
\end{figure*}

\begin{figure*}
\centering
  \includegraphics[width=\textwidth]{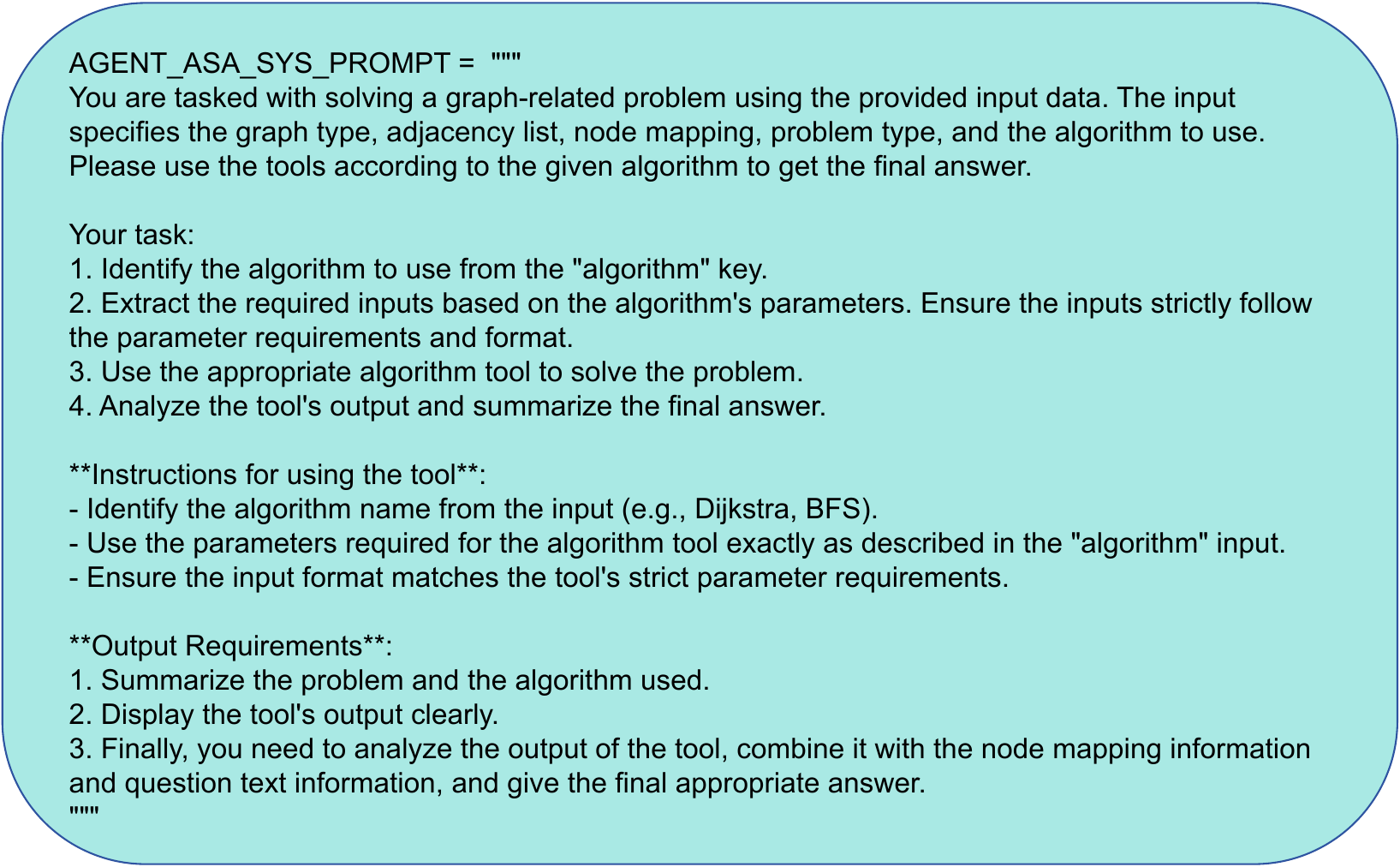}
  \caption {Details of ASA}
    \label{fig:11} 
\end{figure*}

In this section, I will introduce the prompts for each agent, which will be displayed in Figures \ref{fig:6} to \ref{fig:11}.
\label{subsec:A3}
\section{Details on baseline models}
\label{sec:B}
We evaluated 6 of the latest LLMs, including OpenAI o3-mini reasoning model, launched on January 31, 2025 and the latest open-source model, DeepSeek-V3. Table \ref{tab:9} presents more details on the models and their versions.
\section{Details on G-REAL}
\label{sec:C}
\begin{figure*}
\centering
  \includegraphics[width=\textwidth]{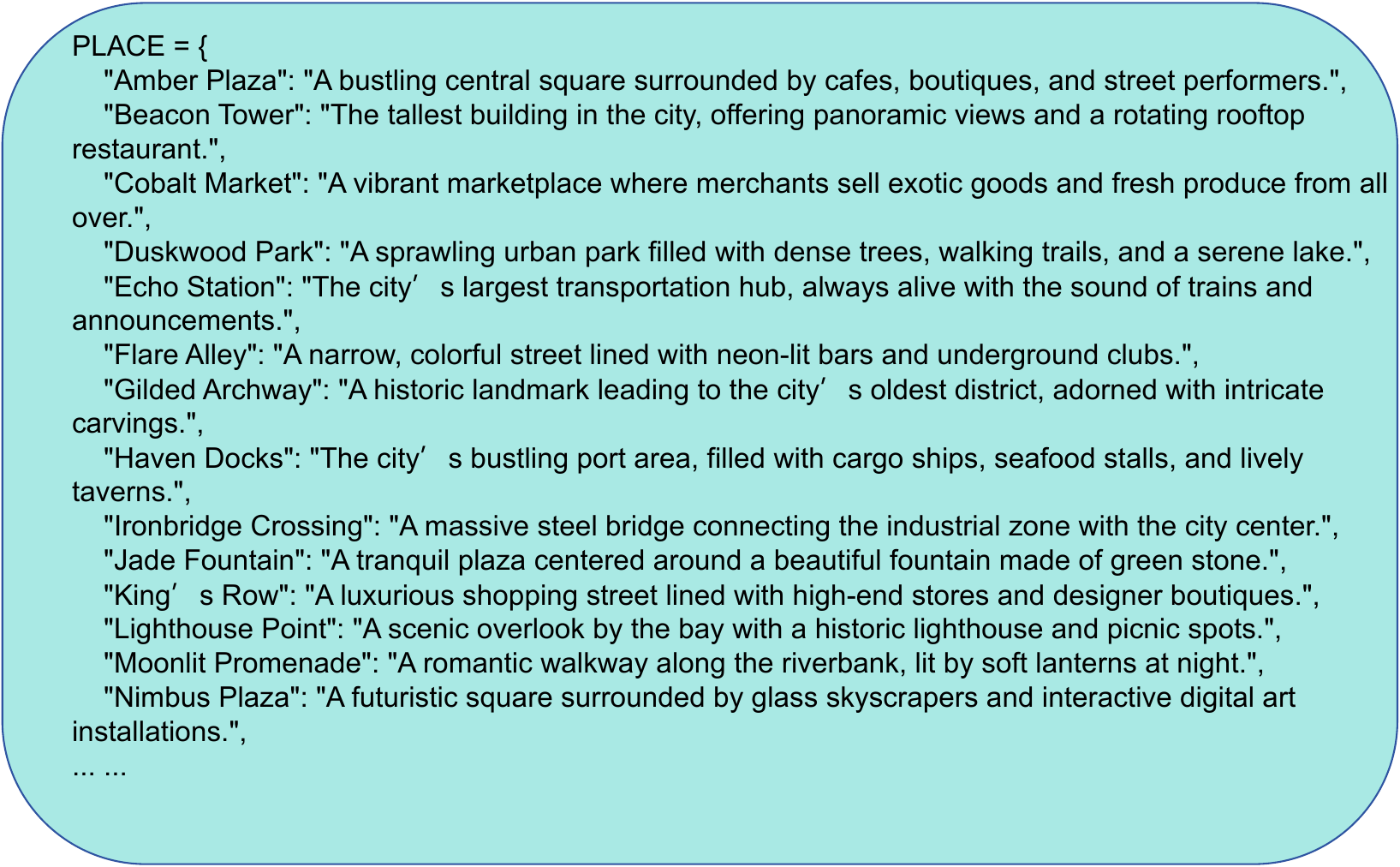}
  \caption {Details of Random Places}
    \label{fig:12} 
\end{figure*}

\begin{figure*}
\centering
  \includegraphics[width=\textwidth]{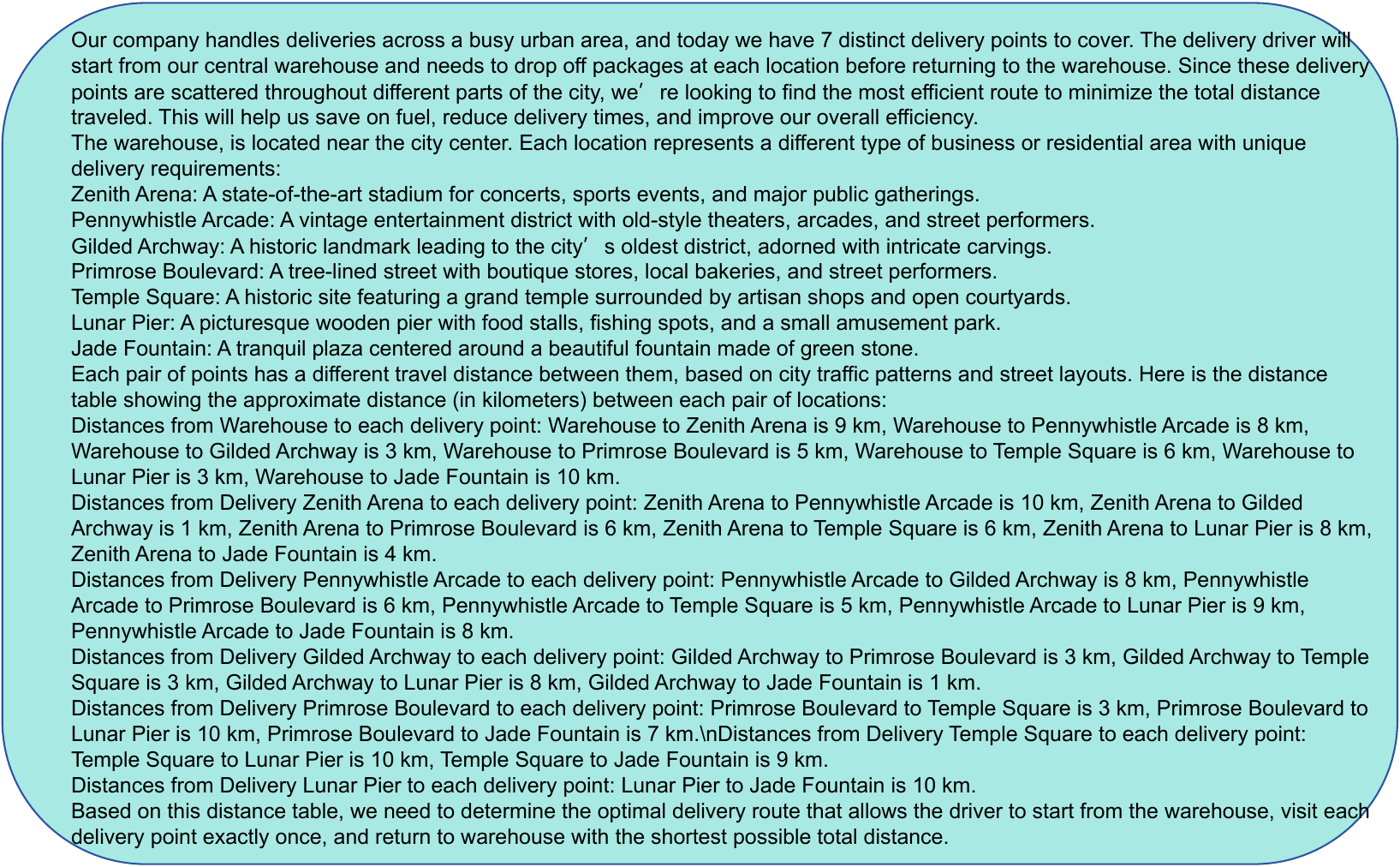}
  \caption {Details of TSP}
    \label{fig:13} 
\end{figure*}

\begin{figure*}
\centering
  \includegraphics[width=\textwidth]{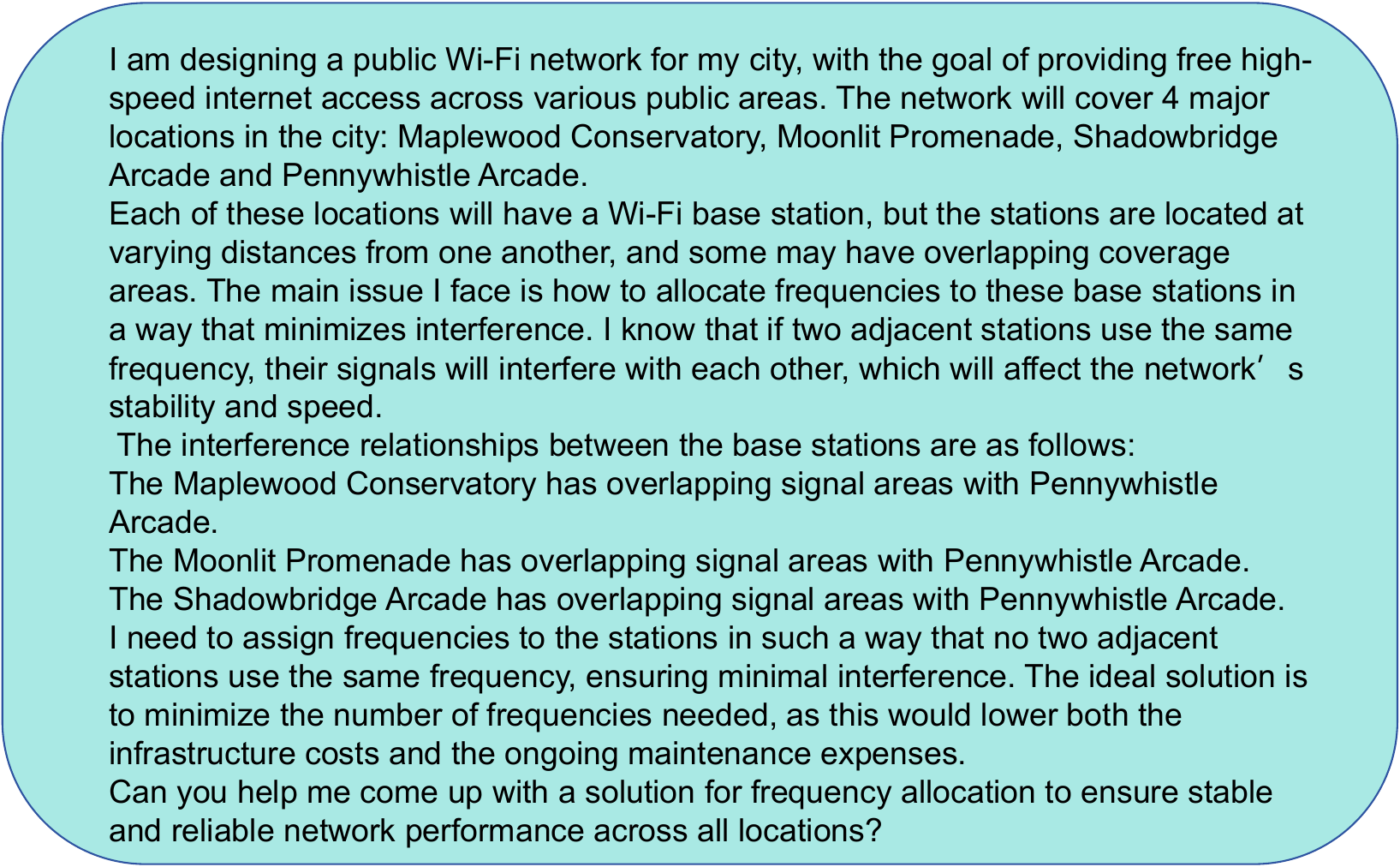}
  \caption {Details of Coloring Problem}
    \label{fig:14} 
\end{figure*}

\begin{figure*}
\centering
  \includegraphics[width=\textwidth]{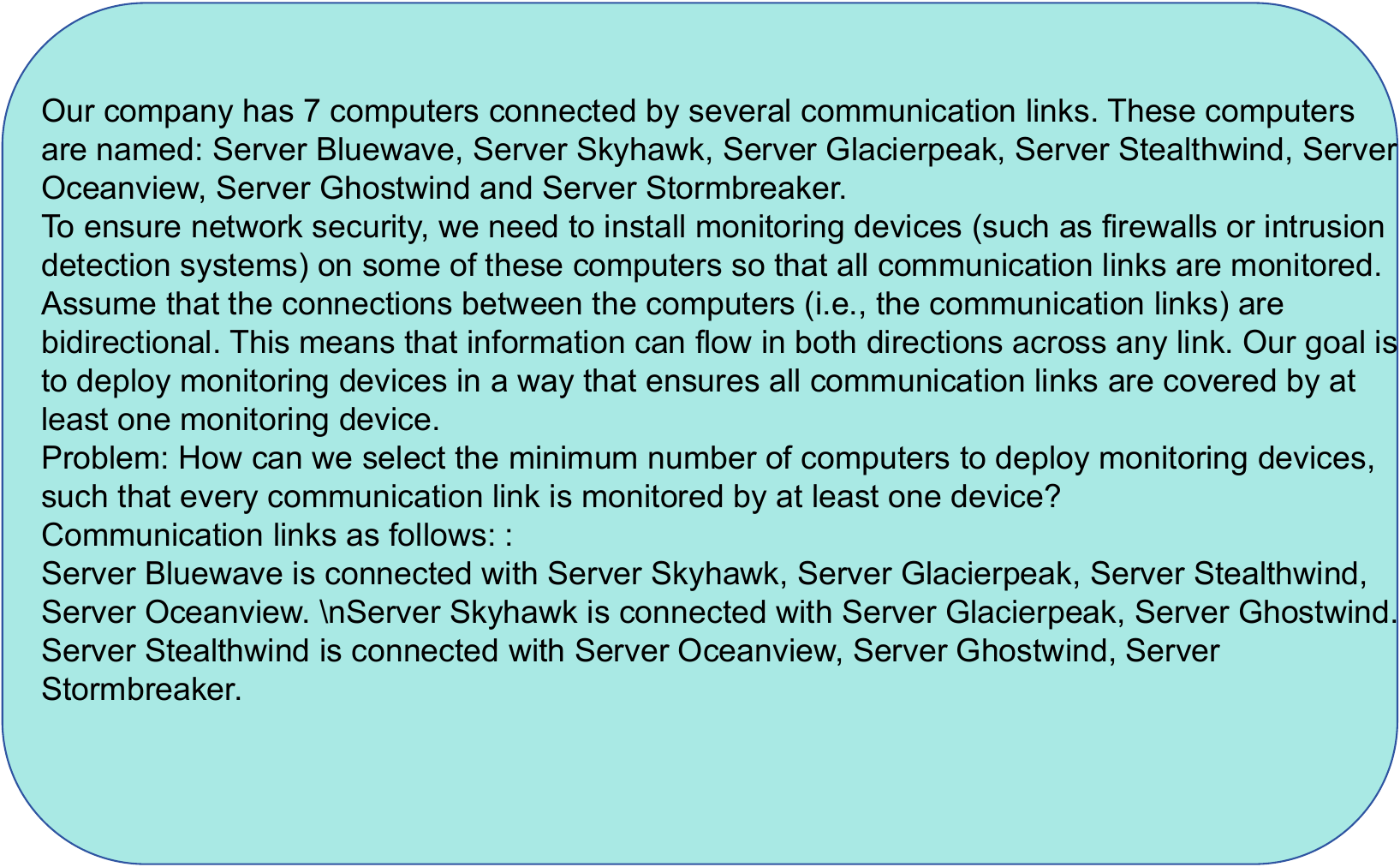}
  \caption {Details of Vertex Cover Problem}
    \label{fig:15} 
\end{figure*}

Existing graph theory benchmarks do not align with real-world scenarios. To better evaluate the ability of MA-GTS in solving graph theory problems in practical contexts and to test the performance gap between LLMs on structured textual graph data and implicit representations, we constructed the G-REAL dataset. This dataset contains three common real-world problems, with detailed information provided in the G-REAL section. It generates problem graphs of varying scales by randomly encoding node names and structures, with the naming conventions and sample problems illustrated in Figures \ref{fig:12} to \ref{fig:15}.
\section{More experimental analysis}
\label{sec:D}
\subsection{Large-scale node analysis}
\textbf{Why 8-25 enough:} In our work, we chose to focus on graphs with 8–25 nodes, primarily due to the complexity and reasoning difficulty posed by realistic tasks. Unlike large-scale but structurally explicit graphs, the G-REAL dataset introduces substantial textual noise, implicit graph structures, and randomly named nodes. These factors make the problem setting significantly closer to real-world semantic reasoning scenarios and increase the overall problem difficulty. This differs from existing benchmarks, which typically construct graph-structured data using explicitly defined and concise textual descriptions.

\noindent\textbf{More experiments:} We have extended our experiments to include larger graph sizes. As shown in Table \ref{tab:6}, we tested the TSP and Graph Coloring problems with 25, 30, 35, and 40-node graphs, with 5 instances evaluated for each size. The results show that even on larger graphs, our framework maintains high accuracy and stability. We plan to include more experiments on even larger graph sizes in future versions of the paper to further validate the scalability of our approach.
\begin{table}[h]
\centering
\begin{adjustbox}{max width=\linewidth} 
\begin{tabular}{c|cccc}
\hline
\textbf{}         & \textbf{25} & \textbf{30} & \textbf{35} & \textbf{40} \\ \hline
\textbf{TSP}      & 0.8         & 0.6         & 0.8         & 0.8         \\ \hline
\textbf{Coloring} & 1           & 0.6         & 1           & 0.6         \\ \hline
\end{tabular}
\end{adjustbox}
     \caption{This table shows the results of experiments conducted on TSP and Coloring problems with extended graph sizes of 25, 30, 35, and 40 nodes. For each size, 5 problem instances were tested.}
    \label{tab:6}
\end{table}

\subsection{Runtime analysis}
G-REAL focuses on graph reasoning under complex semantic conditions, which more closely resemble real-world user scenarios. These tasks often contain intricate semantic information and irrelevant noise, posing significant challenges for LLMs. For instance, the Delivery Logistics Problem in G-REAL is a TSP instance. In contrast to existing TSP benchmarks—where nodes are ordered, connections are explicitly stated, and the problem type is clearly defined—G-REAL requires the model to infer all of this information from natural language. This increases the difficulty of graph construction and makes reasoning more error.

We conducted a supplementary evaluation on the G-REAL-TSP task by randomly selecting five graphs with 15-node scales (Table \ref{tab:7}). We measured the average solution time per problem instance, including task decomposition, tool invocation, and result verification. Compared to existing multi-agent frameworks, MA-GTS demonstrates a clear advantage in time efficiency. These results highlight that our framework is capable of maintaining high accuracy while keeping inference time relatively low, further validating its practical applicability.

\begin{table}[h]
\centering
\begin{adjustbox}{max width=\linewidth} 
\begin{tabular}{c|ccc}
\hline
\textbf{}             & \textbf{MA-GTS} & \textbf{Graphteam} & \textbf{OWL} \\ \hline
\textbf{Time use (s)} & 148.48          & 251.34             & 139.39       \\ \hline
\textbf{ACC (\%)}     & 100             & 0                  & 0            \\ \hline
\end{tabular}
\end{adjustbox}
     \caption{The table presents the results of testing on the G-Real-TSP problem using 5 randomly selected graphs with 15 nodes each.Base model is GPT-4o-mini.}
    \label{tab:7}
\end{table}
\section{G-REAL Details}
\label{sec:E}

G-REAL introduces two main innovations over existing graph reasoning benchmarks. First, it situates graph problems in realistic task contexts with background noise, where node names are randomized and both node and edge descriptions contain distracting language. This design substantially increases the difficulty of structure extraction and semantic understanding, in contrast to benchmarks such as GraphArena that rely on clean, explicit inputs. Second, G-REAL effectively mitigates data contamination by ensuring that randomized identifiers and artificial noise prevent test samples from matching pretraining corpora. As summarized in Table \ref{tab:8}, these features make G-REAL more representative of real-world tasks, more robust against leakage, and more focused on challenging semantic reasoning—aligning it closely with the practical difficulties faced by LLMs.

\begin{table*}[htbp]
    \centering
    \small
    \resizebox{\linewidth}{!}{

    \begin{tabular}{
>{\columncolor[HTML]{EFEFEF}}c |ccccc}
\hline
{\color[HTML]{333333} \textbf{}}                         & \cellcolor[HTML]{EFEFEF}{\color[HTML]{333333} \textbf{G-REAL}} & \cellcolor[HTML]{EFEFEF}{\color[HTML]{333333} \textbf{GraCoRe}} & \cellcolor[HTML]{EFEFEF}{\color[HTML]{333333} \textbf{NLGraph}} & \cellcolor[HTML]{EFEFEF}{\color[HTML]{333333} \textbf{GraphArena}} & \cellcolor[HTML]{EFEFEF}{\color[HTML]{333333} \textbf{GraphInstruct}} \\ \hline
{\color[HTML]{2C3A4A} \textbf{Graph Size}}               & 8-25                                                           & 8-30                                                            & 5-35                                                            & 5-30                                                               & 5-35                                                                  \\ \hline
{\color[HTML]{2C3A4A} \textbf{Real-World Problem}}       & \CheckmarkBold                                                           & \XSolidBold                                                           & \XSolidBold                                                           & \CheckmarkBold                                                               & \XSolidBold                                                                 \\ \hline
{\color[HTML]{2C3A4A} \textbf{Text Noise}}               & \CheckmarkBold                                                           & \XSolidBold                                                           & \XSolidBold                                                           & \XSolidBold                                                              & \XSolidBold                                                                 \\ \hline
{\color[HTML]{2C3A4A} \textbf{Random Naming}}            & \CheckmarkBold                                                           & \XSolidBold                                                           & \XSolidBold                                                           & \XSolidBold                                                              & \XSolidBold                                                                 \\ \hline
{\color[HTML]{2C3A4A} \textbf{Avoid Data Contamination}} & \CheckmarkBold                                                           & \XSolidBold                                                           & \XSolidBold                                                           & \XSolidBold                                                              & \XSolidBold                                                                 \\ \hline
{\color[HTML]{2C3A4A} \textbf{Generation Method}}        & Random Generation                                              & Random Generation                                               & Random Generation                                               & Knowledge graph extraction                                         & Random Generation                                                     \\ \hline
\end{tabular}
}
    \caption{Comparison with mainstream benchmarks.}
    \label{tab:8}
\end{table*}
\begin{table*}[htbp]
    \centering
    \small
    \resizebox{\linewidth}{!}{

    \begin{tabular}{ccc}
        \toprule
        \textbf{Model}           & \textbf{Version}           & \textbf{Model Link} \\ 
        \midrule
        \textit{OpenAI o3-mini}       & o3-mini                    & \url{https://platform.openai.com/docs/models/o1\#o3-mini} \\
        \textit{GPT-4o-mini}          & gpt-4o-mini                     & \url{https://platform.openai.com/docs/models/gpt-4o-mini} \\

        \textit{GPT-3.5}         & gpt-3.5-turbo              & \url{https://platform.openai.com/docs/models/gpt-3-5-turbo}\\

        \textit{Llama3-ins-8b}   & Meta-Llama-3-8B-Instruct   & \url{https://huggingface.co/meta-llama/Meta-Llama-3-8B-Instruct}\\

        \textit{Qwen2.5-7b-ins}    & Qwen2.5-7B-Instruct          & \url{https://huggingface.co/Qwen/Qwen2-7B-Instruct}\\
        \textit{Deepseek-V3}    & DeepSeek-V3-0324-660B         & \url{https://huggingface.co/deepseek-ai/DeepSeek-V3-0324}\\
        \bottomrule
    \end{tabular}}
    \caption{More details about models.}
    \label{tab:9}
\end{table*}
\end{document}